*Review*

# A Survey of NOMA: State of the Art, Key Techniques, Open Challenges, Security Issues and Future Trends

**Syed Agha Hassnain Mohsan** [1,*], **Yanlong Li** [1,2]

1. Optical Communications Laboratory, Ocean College, Zhejiang University, Zheda Road 1, Zhoushan 316021, China; hassnainagha@zju.edu.cn (S.A.H.M.); lylong@zju.edu.cn (Y.L.)
2. Ministry of Education Key Laboratory of Cognitive Radio and Information Processing, Guilin University of Electronic Technology, Guilin 541004, China

* Correspondence: hassnainagha@zju.edu.cn

**Abstract:** Non-orthogonal multiple access (NOMA) systems can serve multiple users in contrast to orthogonal multiple-access (OMA), which makes use of the limited time or frequency domain resources. It can help to address the unprecedented technological advancements of the sixth generation (6G) network, which include high spectral efficiency, high flexibility, low transmission latency, massive connectivity, higher cell-edge throughput, and user fairness. NOMA has gained widespread recognition as a viable technology for future wireless networks. The main characteristic that sets NOMA apart from the conventional orthogonal multiple access (OMA) techniques is its ability to handle more users than orthogonal resource slots. NOMA techniques can serve multiple users in the same resource block by multiplexing users in power or code domain. The purpose of this paper is to provide a thorough overview of the promising NOMA systems. Initially, we discuss the state-of-the-art and existing literature on NOMA systems. This study also examines the practical deployment of NOMA implementation and key performance indicators. An overview of the most recent NOMA advancements and applications is also given in this survey. We also briefly discuss that multiple-input multiple-output (MIMO), visible light communications, cognitive and cooperative communications, intelligent reflecting surfaces (IRS), unmanned aerial vehicles (UAV), HetNets, backscatter communication, mobile edge computing (MEC), deep learning (DL), and other emerging and existing wireless technologies can all be flexibly combined with NOMA. This study surveys a thorough analysis of the interactions between NOMA and the aforementioned technologies. Lastly, we will highlight a number of difficult open problems and security issues that need to be resolved for NOMA, along with pertinent possibilities and potential future research directions.

## 1. Introduction

Demanding applications like high-definition (HD) movies, online gaming, and virtual reality (VR) have caused an information expansion over the past decade [1]. The current challenges presented by the introduction of fifth-generation (5G) include those related to widespread connection, high data rate, ultra-reliability, reduced latency, and energy efficiency, spectral efficiency [2]. Massive heterogeneous networks built on the Internet of Things (IoT) have grown quickly, which forced the development of 5G technology along with introducing a number of new challenges. Because of the uplink or downlink movement of significant user data between various networks, there exist significant barriers to the adoption of emerging multiple access techniques. Due to the aforementioned issues, NOMA, an intriguing and probable 5G network solution, has recently gained a lot of attention [3]. Both the business sector and academic fraternity have identified NOMA as a technological trend that holds promise for addressing the various needs of 5G. This essential enabling technology is necessary for next-generation wireless or mobile networks to meet the various requirements for throughput, fairness, data rate, redundancy, and connectivity. NOMA can integrate multiple users in each resource block with the incorporation of signal superposition. In order to increase throughput, NOMA techniques can also provide power allocation to nodes with weak channel conditions. NOMA has the potential to effectively use the available resources than conventional multiple access (MA) techniques [4, 5]. In contrast to NOMA, conventional MA approaches serve a single user with each orthogonal resource unit. However, it degrades the spectrum efficiency and overall throughput of the system. When NOMA is utilized in such situations, it ensures that users with better channel conditions can also consume the same amount of bandwidth as users with weaker channel characteristics. After multiplexing at the transmission side using superposition coding (SC) at different power levels, the multiplexed NOMA signals are broadcasted to the end users. Users with poor channel characteristics frequently receive high power than those with superior channel characteristics. As a result, with NOMA, it is now more crucial to send data using the correct channel state information (CSI). Users can easily retrieve the signals that have accumulated in the receiver with a strong channel gain. Weak channel gain causes users to perceive other signals for interference, which drastically reduces spectral efficiency. This critical issue might be overcome with NOMA without guard period and no signal interference [6].

5G may be able to offer significant faster data speeds, more connections from mobile devices, a larger system's capacity, lower latency, and less power usage. The 5G-aided systems, including IoT systems, are made up of a wide range of data transfer entities with different demands for data rate and latency. The main purposes of NOMA in 5G systems are to achieve high spectrum efficiency, increase user density, and ensure low latency [7]. NOMA is used in 5G systems where multiple users rely on the same resources. NOMA in 5G systems is expected to meet the requirements of high spectrum efficiency and reliable connectivity among numerous data transfer entities. However, NOMA systems have particular shortcomings, including high computational complexity, challenging designs, and resource allocation concerns. Additionally, NOMA systems are prone to perfect CSI in order to carry out sequential interference cancellation at the receiver end. For enhanced performance of NOMA, consecutive interference nullification must be perfect. It is generally difficult to develop a good power allocation (PA) approach without knowing the right CSI at the transmitter side. Therefore, obtaining an optimal-perfect or perfect CSI could be difficult to achieve. The use of deep learning (DL) technology can help to overcome these limitations. DL techniques mainly improve the performance of various wireless communication systems. For a wide range of applications, current communication systems mainly rely on DL-assisted NOMA technology. In a recently reported work [8], the authors provide a comprehensive review of DL techniques for NOMA systems. They briefly discuss several challenges and future research aspects of DL-based NOMA systems. Figure 1 presents a general illustration of downlink and uplink

NOMA for 2 users. Table 1 summarized several NOMA techniques towards 5G standardization discussed in literature.

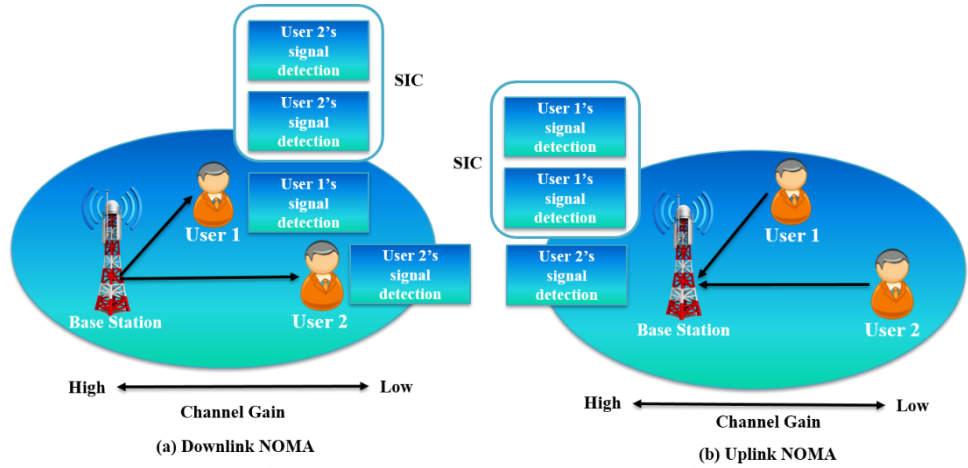

**Figure 1.** An illustration of downlink and uplink NOMA

**Table 1.** Summary of NOMA techniques towards 5G standardization [9,10]

| NOMA scheme | Technical point | Key benefits |
|---|---|---|
| RSMA | Low PAPR modulation | Extended coverage |
| RDMA | CP-based time-frequency repetition | Ease of implementation |
| NOCA | Zadoff-Chu sequence | Low PAPR, Easy to generate |
| GOCA | Grouping for orthogonal or non-orthogonal sequences | Inter-group orthogonality, group separation |
| MUSA | Short-complex spreading sequence | Large number, Easy to create |
| PDMA | Irregular LDS | Irregular protection |
| SCMA | Multi-dimensional modulation | Signal space diversity gain, more diversity, constellation shaping gain, without CSI |
| LCRS | Bit level spreading, Low rate FEC code | Large coding gain |
| LSSA | Interleaving, Low rate FEC code | Large number of signatures |
| IDMA | Low FEC code, Bit-level interleaving | Randomized the mutual interference |
| BOMA | User multiplexing | Additional flexibility in selecting component codes |
| LPMA | User multi-level lattice superposition | Enhanced diversity and flexibility |
| SDMA | User-specific channel impulse response | To support large number of users |
| IGMA | Low coding rate, Bit-level interleaving | Sparse grid mapping, easy user separation |
| NCMA | Grassmannian line packaging algorithm | New generations of codebooks satisfy the Welch bound with equality. |
| FDS | Multiple orthogonal or quasi-orthogonal codes | Directly spreads the modulation symbols |
| SAMA | Spreading sequence with variable sparsity | More diversity by spreading users data over various resources |
| PD-NOMA | Users transfer data with various power levels considering same RB | SIC, higher system level performance than OMA, less receiver complexity |
| LDS-CDMA | Users share an RB through user-specific spreading sequence | Limit interference on each chip |
| LDS-OFDM | Symbols are multiplied with LDS | More feasible for wideband than LDS-CDMA |
|  | sequence and mapped onto OFDM subcarriers |  |
| LDS | Spreading codes | Without CSI |
| PDMA | Pattern based multiplexing | Enhanced performance, improved average sum rate of users |

*1.1. Comparison between NOMA and OMA*

OMA schemes (OFDMA, CDMA, FDMA, and TDMA) multiplex signals from various users on orthogonal resources either in code, frequency, or time domain. In contrast to OMA, NOMA multiplexes signals of various users in code or power domain while sharing the same code, frequency or time resource block (RB) providing high bandwidth efficiency [11-12]. In OMA, multiplexing gain and good throughput can be attained through simple cost-efficient receivers which can mitigate inter-user interference. However, traditional OMA schemes are limited by expanding number of users linked through the orthogonal resources. Moreover, due to limited BW resource allocation, these scheme have low spectral efficiency. In contrast, NOMA offers higher spectral efficiency, improved quality-of-service (QoS), enhances quality of user experience (QoE), low transmission latency, higher capacity, massive connectivity, high transmission rate, fairness trade-off and good compatibility with various access techniques.

NOMA supports user fairness while achieving minimum required rate and enhances spectral efficiency through upgraded receiver designs to nullify inter-cell and intra-cell interference. The main idea in NOMA is power allocation where high power is allocated to the user with weak channel gain and less power is allocated to the user with strong channel gain. This indicates a balance between user fairness and overall system throughput. NOMA has the potential to serve multiple users over the same RB at the cost of inter-user interference. NOMA exploits received power disparity or a specific kind of complex receiver termed as SIC [12] to retrieve the desired signals and mitigate inter-user interference. SIC is implemented at the BS in case of uplink NOMA or at the strong user side in case of downlink NOMA. It supports to decode signals of the multiplexed users and improve channel capacity. The enhanced receiver complexity with expanding number of users at the need for unique channel gain difference of multiplexed users for better performance in PD-NOMA can be overcome through hybrid OMA-NOMA [13,14]. NOMA supports the simultaneous transfer of multiplexed users through same degree of freedom (DoF) using superposition coding with different power levels. It incorporates multiple user detection (MUD) for the separation of multiplexed users sharing the same DoF as shown in Figure 2. NOMA has the capability to support more number of users in the same DoF through controllable symbol collision. The flexibility of NOMA compared to current techniques makes it compatibility to integrate to achieve hybrid variant such as hybrid OMA-NOMA. Table 2 summarizes comparison between NOMA and OMA schemes.

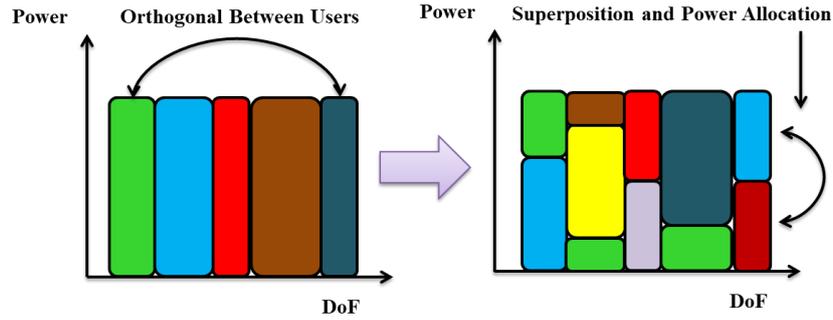

**Figure 2.** From OMA to NOMA through power domain multiplexing.

**Table 2.** Advantages and disadvantages of OMA and NOMA

|  | **OMA** | **NOMA** |
|---|---|---|
| **Advantages** | ✓ Low receiver complexity | ✓ Low signaling and latency<br>✓ High compatibility<br>✓ Massive connectivity<br>✓ Enhanced user fairness<br>✓ Increased spectral efficiency |
| **Disadvantages** | ✓ High transmission latency<br>✓ Limited number of users<br>✓ Bad users fairness<br>✓ Reduced spectral efficiency | ✓ Increases the inter-cell interference<br>✓ Increases the signaling overhead<br>✓ High receiver complexity<br>✓ Increases the CSI feedback overhead |

*1.2. Scope and Contributions*

Readers who have a deep interest to read emerging NOMA systems must go through this review. This article surveys an in-depth analysis of existing NOMA systems in literature. The key motive behind this survey is to help researchers with a thorough understanding of the NOMA systems from an extensive range of perspectives. It offers a brief discussion of NOMA's benefits, key performance indicators and promising applications. It also discusses the coexistence of NOMA with several emerging technologies. Finally, it highlights several potential challenges, security issues and future trends for further breakthrough novelties.

*1.3. Organization of the Paper*

We have organized this survey as follows. Section 2 discusses the existing studies on NOMA systems. Section 3 discusses the key performance indicators for NOMA techniques. While Section 4 discusses the integration of several emerging technologies with NOMA. Section 5 outlines various applications of NOMA techniques. Section 6 investigates potential challenges and open research problems. Section 7 investigates some security issues of NOMA. Section 8 is dedicated to future trends. Lastly, Section 9 brings the paper to an end.

**2. Related Surveys**

The authors of [5] examined the fundamentals of NOMA and compared it with OMA while considering the spectrum efficiency, complexity, and performance of power domain-NOMA (PD-NOMA) and code domain-NOMA (CD-NOMA). The authors of [16] discussed several user pairing strategies, power allocation methods, and real-world implementation problems for downlink PD-NOMA. Multiple-input and multiple-output (MIMO) non orthogonal multiple access (MIMO-NOMA) and MIMO orthogonal multiple access (MIMO-OMA) performance comparisons are also conducted. The researchers in [17] have thoroughly examined the fundamental principles, most recent developments, and future research trends of NOMA. Particularly, the fundamental principles of NOMA and specifically comparison with OMA considering the information theory perspective are discussed. Moreover, they compared the spectrum efficiency, system performance, and receiver complexity of NOMA system designs. They also highlighted a variety of open problems, associated solutions, and future research areas to deal with these crucial concerns. A flexible incorporation of NOMA with several emerging techniques such as MIMO, mmWave, cognitive radio (CR), physical layer security (PLS), energy harvesting, VLC, and mobile edge computing (MEC) has been described in [18]. With the aid of promising technologies like MIMO, CR, mMIMO, VLC, mmWave, MEC, unmanned aerial vehicles (UAV), and underwater wireless communication (UWC), the rate-optimal PD-NOMA schemes are investigated in [19]. The authors in [20] have investigated traditional NOMA misconceptions and the benefits of incorporating NOMA with machine learning (ML) and deep learning (DL) techniques. The authors of [21] implemented NOMA to substantially increase spectrum efficiency. They also examined the NOMA methods for spectrum sharing in 5G networks. They focused on NOMA's research challenges and critical issues for 5G scenario. The authors of [22] carried out simulation analysis for Welch-bound equality spread multiple access (WSMA)-based NOMA and multiuser-MIMO (MU-MIMO). They discussed about the 3GPP standard's NOMA and provided a number of methods to lessen the complexity and delay of downlink and uplink NOMA. The primary challenges with resource distribution, practical applications, signaling concerns, security methods, and NOMA restrictions were discussed in [1]. Recent studies focus on optimization, matching theory, game theory, ML approaches, and graph theory [23]. In an explicit overview of PD-NOMA-based VLC systems, some limitations and technical challenges including the clipping effect, power allocation, and security concerns are highlighted [24]. In the context of VLC-based NOMA systems, challenges and promising future research avenues are also identified [25]. The authors also investigated deployment of PD-NOMA-based VLC systems, open research problems and their mitigative solutions. In [26], the authors have discussed PD-NOMA in the context of cooperative networks. In [27], the authors have focused on deep learning NOMA to support 5G networks. In another recent study [8], the authors have provided a comprehensive review of DL-based NOMA systems. They investigated key performance indicators, integration with several promising technologies, potential challenges, and future research trends. Table 3 summarized several recent studies reported in literature on NOMA systems.

**Table 3.** Summary of existing studies on NOMA

| Reference | Year | Key focus |
|---|---|---|
| [13] | 2017 | This study provides a comprehensive survey of NOMA for 5G technologies. It discusses several challenging issues and future trends. |
| [15] | 2018 | It focuses on CD-NOMA, PD-NOMA, and their comparison regarding spectral efficiency and complexity. |
| [16] | 2018 | It discusses user pairing, power allocation and challenges for downlink PD-NOMA. It also compares MIMO-OMA and MIMO-NOMA. |
| [17] | 2018 | A comprehensive survey on NOMA, recent developments, challenges, associates solutions, and future research trends. It provides systematic comparison of CD-NOMA and PF-NOMA considering system performance, spectral efficiency and receiver complexity. |
| [18] | 2019 | It focuses on the incorporation of NOMA with various emerging technologies including MEC, VLC, security, energy harvesting, CC, CR, mMIMO, and mmWave etc. |
| [19] | 2019 | It discusses PD-NOMA and existing technologies such as VLC, mmWave, MEC, CR, MIMO, mMIMO, UWC, UAV, and CC in the perspective of PD-NOMA. |
| [20] | 2019 | It investigates NOMA in 5G scenario and role of ML and DL approaches. |
| [21] | 2019 | It discusses NOMA-aided survey for spectral sharing towards 5G networks. It |

| | | also highlights several challenges and research issues of NOMA in 5G networks. |
|---|---|---|
| [22] | 2020 | It surveys simulation comparison of MU-MIMO and WSMA-enabled NOMA. It also discusses different techniques to overcome the deployment problem and delay of uplink/downlink NOMA. |
| [1] | 2021 | It focuses on practical implementation, security issues, resource allocation, and signaling aspects in NOMA. It also addresses challenges and future aspects of PD-NOMA towards 5G wireless networks. |
| [23] | 2022 | It addresses NOMA in the perspective of 5G/B5G networks. It also examines several NOMA challenges and related solutions. It also discusses MM, EN, FN, CV, Blockchain and its role in healthcare. |
| [24] | 2022 | It overviews PD-NOMA aided VLC systems. It also highlights security, power allocation, clipping effect, research issues and future research directions. |
| [8] | 2023 | It provides a comprehensive review of DL-based NOMA techniques. It discusses the role of DL approaches in NOMA, security issues, practical challenges and future research directions. |
| [25] | 2023 | It surveys VLC-based NOMA techniques. It highlight integration of VLC-NOMA with several emerging technologies, practical challenges, and future research directions. |

*2.1 NOMA's Benefits*

The NOMA technology offers several appealing benefits which will be covered here.

• Massive Connectivity

It seems like research fraternity has shown an agreement that NOMA is necessary for large connectivity. This is due to the fact that all traditional OMA approaches have an inherent resource constraint on the number of supplied users (such as the number of resource blocks and codes in OFDMA and CDMA, respectively). In contrast, NOMA theoretically service any number of users by superimposing their signals even inside a single resource block. In this way, NOMA can be adapted to IoT use cases where a lot of interconnected entities intermittently broadcast a limited number of data packets. In fact, it is severely inefficient to provide one device access to a full resource block which was previously implemented in OMA [28].

• Low Latency

The latency needs for 5G use cases usually vary according to the application scenario. For instance, for improved mobile broadband (eMBB) and ultra-reliable and low-latency communications (URLLC), the ITU mandates user plan latency requirements of 4 ms and 1 ms, respectively [29]. Because a device has to wait until the availability of an empty resource block, no matter how many bits it intends to send, OMA makes it crucially hard to guarantee such strict latency requirements. Contrarily, NOMA encourages flexible scheduling because it can handle multiple devices, resulting into an improved QoS.

• High Spectral Efficiency

IMT-2020 specifications state that the downlink's maximum spectral efficiency must be 30 bits/s/Hz. In comparison to OMA, NOMA delivers a better user-fairness and higher spectral efficiency. NOMA features the theoretically ideal method of exploiting the spectrum for both downlink and uplink communications in a single-cell network. NOMA can achieve an enhanced performance as each NOMA user has access to the complete bandwidth as compared to OMA where users are only allowed to use a limited spectrum. To further ensure higher throughput and enhance spectral efficiency, NOMA can also be integrated with other cutting-edge technologies, such as VLC, MEC, UAVs, millimeter wave (mmWave) and massive MIMO technologies [30-31].

**3. NOMA's Key Performance Indicators (KPIs)**

This Section focuses on some key performance indicators which must be kept into account before NOMA can be properly implemented in real-world wireless communication systems.

*3.1 Modulation and Coding for NOMA*

For NOMA, efficient modulation and channel coding and methods are essential to achieve the theoretically expected rates in practical applications. For instance, pulse amplitude modulation (PAM) is used with turbo codes, gray labeling, and NOMA in [32] and [33]. It is demonstrated that the promising NOMA scheme is superior to traditional OMA. Other channel codes, aside than turbo codes, are also used with NOMA in [34]. The authors in [35] investigated the effect of finite-alphabet inputs on NOMA aided Z-channels. More significantly, the combination of modulation and complex coding with NOMA has produced new variations of the technology, including LPMA [34] and Network-Coded Multiple Access (NCMA) [36]. Figure 3 shows an LPMA downlink scenario with two users. LPMA uses lattice coding to encode two users' messages. The two encoded messages which are multiplied with a prime number are linearly combined to as a transmitted signal. For instance, the weak user's message and strong user's message are multiplied by a larger and smaller prime number, denoted by $p_1$ and $p_2$, respectively. At the receiver, the modulo operation is used to eliminate multiple access interference. To cancel out the strong user's message, the weak user applies a modulo operator with regard to $p_2$. We see that the way LPMA eliminates interference from multiple access is remarkably similar to direct-sequence CDMA (DS-CDMA). Since the chip rate of CDMA is significantly higher than the data rate, LPMA significantly eliminates this serious drawback. As demonstrated in [34], LPMA can perform better than traditional PD-NOMA, specifically when both users have similar channel characteristics.

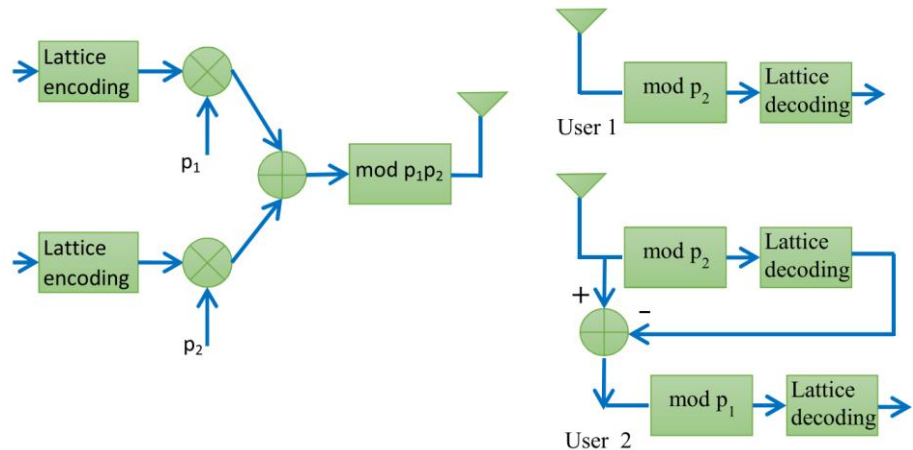

**Figure 3.** Two users based LPMA downlink transmission [13].

*3.2 Cross-Layer Resource Allocation*

Because various users must be accommodated in any NOMA system which makes it a complicated system with associated degrees of freedom for resource allocation, including subcarrier allocation, user grouping/clustering, precoding/beamforming design, and power allocation. Even though resource allocation in a centralized manner can produce the best performance, still the complexity and signaling overhead associated with this strategy must be taken into account [37] and [38]. Distributed resource allocation in NOMA systems has garnered a lot of interest. The energy efficiency of NOMA is explored in [39], where sub-channel assignments and power allocation are concurrently created using the difference of convex (DC) programming technique. Using queuing theory, a joint power allocation and rate control strategy is suggested in [40]. For both downlink and uplink NOMA, the system throughput maximization problem is formulated in [41], where user clustering and power allocation are optimized. In [42], a similar alternating strategy is also investigated by decoupling the joint design of power allocation and user scheduling using proportional fairness. Matching theory can be used to simulate the user pairing issue in the perspective of NOMA if only two users are served at each sub-channel. Matching theory may successfully simulate the dynamic interactions between users when each user has a priority list that indicates the pairing users. This observation led to the development of several matching theory-enabled user pairing mechanisms for NOMA communication scenarios [43].

*3.3 Capacity*

In [44], the authors discusses a comparison of the capacity of MIMO-OMA and MIMO-NOMA with multi-users cluster. In MIMO-OMA and MIMO-NOMA the sum rate and ergodic sum-rate changes are examined against the transmit power for two and three users, respectively. The simulation findings demonstrate that the total rate of both 2-user and 3-user cases in MIMO-OMA and MIMO-NOMA increases by increasing transmit power. Thus, the ergodic operation lowers fluctuations brought on the channel variation. Furthermore, in both the 2-user and 3-user cases, MIMO-NOMA performs better than MIMO-OMA in the context of both aforementioned factors. Additionally, it has been found that in MIMONOMA, the 2-user case performs better than the 3-user case because the sum rate drops by increasing the number of incorporated users in a cluster.

*3.4 Coverage*

The coverage analysis of a dense HetNet with C-NOMA is examined in [45]. The coverage probabilities are computed using a dense HetNet with C-NOMA technique. The simulation findings demonstrate that C-NOMA, particularly for remote users in picocell BSs, can significantly improve the coverage performance of NOMA users if the user intensity is not significantly higher than the intensity of all BSs. The C-NOMA system is hard to be implemented because of the lack of numerous void BSs necessary for coordination, hence an increase in user intensity lowers the likelihood of coverage. Because the near user already conducts error-free SIC practically without assistance from the joint transmission, thus C-NOMA has low impact on the near user's coverage probability. The only case the user intensity is lower than the total intensity of BS is when the close user has a weaker coverage than the distant user.

*3.5 Data rate*

The NOMA system is employed in [46] in order to analyze the sum rate. The sum rate in relation to the SBS count is investigated. The simulation findings demonstrate that when the number of SBSs rises, the sum-rate rises as well. It has been noticed that the sum rate initially increases quickly, but as the number of SBSs rises, the improvement slows gradually. It is due to the constrained backhaul capacity, and as additional SBSs are added to the network, the traffic on the backhaul link grows. As a result, adding SBSs results in higher congestion and interference on the backhaul network. Analysis are performed on how maximizing the level of the self-interference controlling capability (SICC) will affect the sum rate. When the SICC value rises, the system sum-rate falls. Self-interference has a greater impact as SBS transmit power is increased.

*3.6 Bit Error Rate (BER)*

The NOMA technique is employed in [47] to investigate the BER performance. Assuming that each user has a power output of 0.5 and 1 watts. The reference user, who is located closer to the BS, has a power of 0.5 watts, whereas the interfering user, who is located farther away, has a power of 1 watt. The interference causes changes in the received power. Surprisingly, the traditional NOMA receiver first detects the reference user before detecting the signals of the other users in descending sequence of power. As a result, the reference user is detected without any interference, and the performance obtained with traditional NOMA is improved. In this design, the interfering user's SIC receiver cannot identify the reference user's signal since it is delivered at a power level of 0.5 watts, which is 3 dB less than the interfering user. Moreover, the reference user's SIC eliminates the interfering signal because its less power in conventional NOMA, which raises the BER.

*3.7 Massive Connectivity and QoS for NOMA*

The primary function of next-generation multiple access (NGMA) in 6G is to ensure extremely high connection. The massive connectivity offered by NGMA should be controlled by multiple QoS criteria—in the context of user experience—which includes data rates as well as reliability and latency. This is because the majority of 6G applications have diverse QoS requirements. Following that, NGMA can help 6G networks to achieve their highest level of spectral efficiency. It is worth noting that for 6G, a high spectral efficiency also aids in energy conservation. For uplink NOMA, semi-grant-free (semi-GF) transmissions are preferred. While downlink NOMA's architecture becomes QoS-oriented in order to meet these performance criteria and provide massive connectivity. It should be noted that repeated SIC iterations result in lengthy delay and substantial outage possibilities. The key issue for downlink NOMA in 6G is overcoming low reliability and excessive latency owing to SIC, especially for the scenario with multiple users. Along with hybrid SIC to increase spectral efficiency, QoS-enabled NOMA must consider connectivity issue while designing QoS-enabled power allocation (Q-PA) QoS-enabled user clustering (Q-UC) [48].

*3.8 Outage*

In NOMA, probability and sum-rate outages are crucial and must be kept into account. The issue of outages can be compared to problems with power distribution, choosing the best decoding sequence, and user grouping. The error probability is strongly restricted by the outage probability. In addition, outage likelihood can be utilized to compute outage capacity. Research into the NOMA outage is still not very advanced. There are few studies that focus on this crucial factor as a KPI. For instance, in [49] where a general MIMO architecture for NOMA uplink and downlink transmission is introduced, outage probability is taken into account as a KPI. Similarly, the author in [50] introduced a NOMA technique with retransmission. Furthermore, the authors in [51,52] used randomly deployed users to evaluate the performance of NOMA in 5G networks. Moreover, the authors in [53] provided outage achievable rate analysis for NOMA with multiple-relay channel, where the achievable rate is regarded as a KPI.

*3.9 Miscellaneous*

There are several other KPIs which require further research, such as latency, time allocation, energy efficiency and average number of handovers. For instance, the authors in [54] considered EE as a KPI, where they enhanced energy efficiency for the MIMO-NOMA systems. The authors in [55] considered average time allocation as a KPI, where the system performance is optimized considering a resource allocation technique. In [56], the authors considered average number of handovers as a KPI, where they presented a power allocation method for VLC-based NOMA system.

*3.10 Successive Interference Cancellation*

The shortcomings of SIC can be overcome through deep learning approaches. Due to weak interference nullification of SIC receiver, the capacity performance degrades. Due to various hardware challenges, decoding and interference cancellation is difficult to achieve in real-world systems. To overcome these challenges, the authors of [57] suggested that NOMA's performance gained can be increased through SIC at the cell-edge users. It is essential to design an effective and easy-to-use SIC receiver. Multi-stage SIC receiver design can be used to lower BER and multipath fading. SIC receiver's performance can be improved through efficient and low complexity power allocation algorithms [58]. Because of signal processing needed for SIC, the complexity of receiver increases with the increase of number of multiplexed users. In [59], the authors introduced a deep neural network (DNN) model to approximate the SIC. In MIMO-NOMA systems, the joint optimization of SIC decoding along with precoding can reduce the total mean square error between decoded signal and desired user's signal.

*3.11 Channel State Information*

Generally, CSI puts a substantial impact on performance of NOMA system. Several research studies have been reported on CSI for NOMA techniques [60]. In a recent study [61], the authors designed a novel linear estimator to enhance the average effective signal-to-interference noise ratio (SINR) of the strong user along with a finite SINR needed for the weak user to investigate CSI. Furthermore, several research groups are focusing on CSI solutions for NOMA as it is a critical issue in traditional methods. The authors in [62] reported power allocation methods and NOMA system's performance was evaluated using imperfect CSI. In addition, the authors in [63] investigated that inaccurate CSI results in decoding error and introduces additional interference with the desired signal. Thus, it is a crucial concern to efficiently address CSI in NOMA systems and novel methods must be proposed to overcome this issue. Previously, several research studies have introduced reliability and sum rate optimization techniques. However, these techniques usually need high computational complexity because of non-linear optimization. These methods are limited to generate a power allocation method against any given CSI. Particularly, the key advantages of NOMA mainly rely on CSI. Thus, various CSI methods have been introduced in literature to further improve the accuracy of channel estimation [64]. Traditional methods have difficulty to track the changes in the channel state because of channel complexity in multi-user systems [65]. Usually, these variations in channel conditions cause disruption in CSI acquisition and performance degradation in NOMA systems.

*3.12 Power Allocation*

One important factor to enhance NOMA system's performance is power allocation to NOMA users when limited resources are available. It is considered that optimum power allocation is NP-hard which makes it expensive and unfeasible to trace all channel conditions in order to investigate the best ideal solution. In this context, researchers have proposed several methods to overcome this issue. Solutions are based on power allocation for downlink SISO-NOMA [66], power allocation for sum-rate maximization [67], and energy-efficient use of resources [68]. Several studies have proposed DL approached as well. In this context, DL approaches have received significant attention due to ongoing technological advancements for efficient power allocation. In [69], the authors introduced a deep reinforcement learning (DRL) technique to perform power allocation for downlink NOMA system based on single BS and multiple users. Similarly, the motive behind [68] is to optimize user's data rate through power allocation. The proposed system offers a sum-rate comparison between Joint Resource Allocation (JRA) with and without DL. The finding indicate that JRL without DL substantially outperformed its counterpart. The authors of [70] proposed a power allocation strategy with deep learning methods for the optimization of system sum rate considering imperfect SIC in downlink NOMA. Generally, it is very hard to find the best ideal method for power allocation in real-time scenario. In [71], the authors introduced a power allocation technique for imperfect SIC to improve sum rate of NOMA system. This method uses deep learning for the prediction of idea power allocation criteria.

**4. Integration of NOMA with Emerging Technologies**

This section presents a brief discussion of the integration of NOMA with various 5G emerging technologies as shown in Figure 4. Our aim is to highlight the advantages of this integration as investigated in the literature.

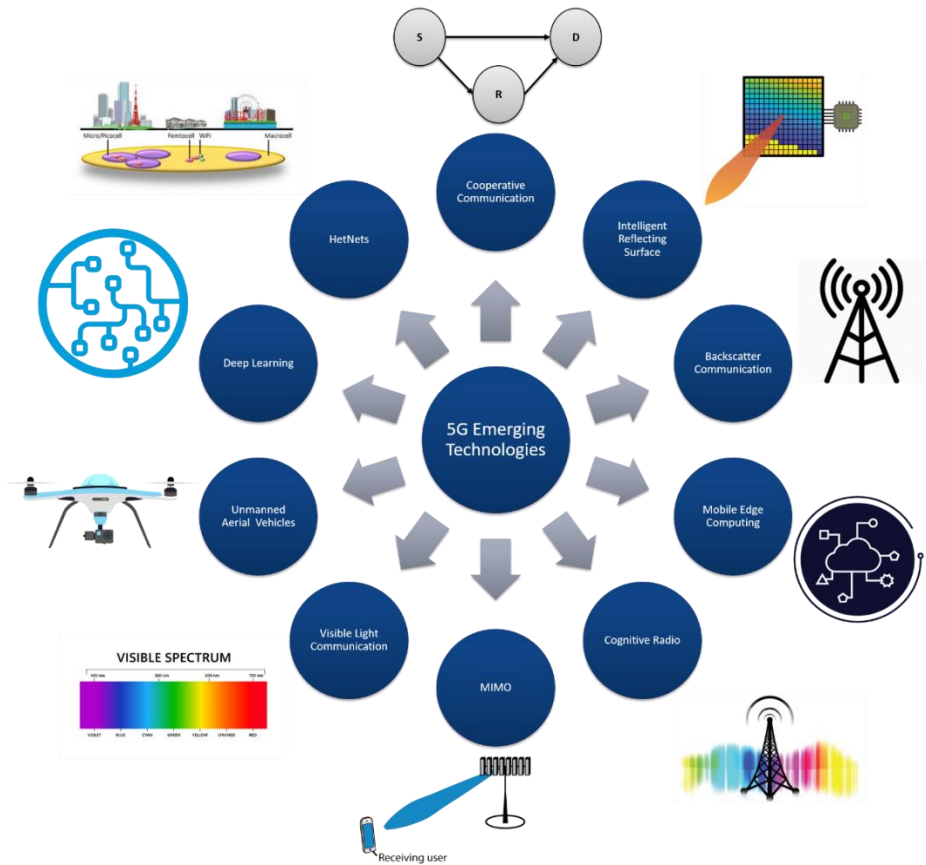

**Figure 4.** 5G emerging technologies.

*4.1. Cooperative NOMA Communications*

Future wireless networks are expected to highly feature different types of collaboration in cooperative NOMA communications as described below. Adoption of rate-optimal techniques under both high and low mobility conditions is a topic that requires exploration for user-aided and relay-aided cooperative NOMA schemes in order to increase spectral efficiency and accommodate more users over scarce spectrum resources. In addition, only a limited number of studies have examined rate-optimal cooperative NOMA systems in downlink configurations, such as [72], whereas further research contributions are required in uplink scenario. Below we have discussed some open research issues [19]:

i) Most of the studies on cooperative NOMA communications have looked into either half or full-duplex mode separately. The possibility for dynamic switching between these modes in user-aided and relay-aided cooperation types has only been investigated in a few publications [73,74]. It is crucial to investigate such novel hybrid mode works with dynamic cooperative systems.

ii) Some researchers have focused on AF and DF relaying protocols in cooperative NOMA systems, while studies of ANC and CF relaying protocols have only been conducted in a few NOMA configurations. Notably, the authors in [75] focused on under relay broadcast channel and the researchers in [76] examined an overlay cooperative-cognitive radio network while considering CF relaying protocol with the NOMA scheme. In order to assist the research fraternity to investigate the best relaying protocol for each application, it is helpful to examine ANC, CF, and maybe additional relaying protocols for various scenarios.

iii) For in-band FD relaying, different SI mitigation technologies were thoroughly studied in [77] and [78]. Future studies should focus on comparing the rate performance of in-band FD-based NOMA systems using various SI mitigation techniques.

iv) A full-duplex user-aided cooperative NOMA system with imperfect and perfect and imperfect SCI n has been studied in [79]. Relay-aided cooperative NOMA systems for various frameworks can be used in such situations.

Table 4 provides a summary of existing studies on CC and NOMA. An illustration of cooperative NOMA communication is presented in Figure 5.

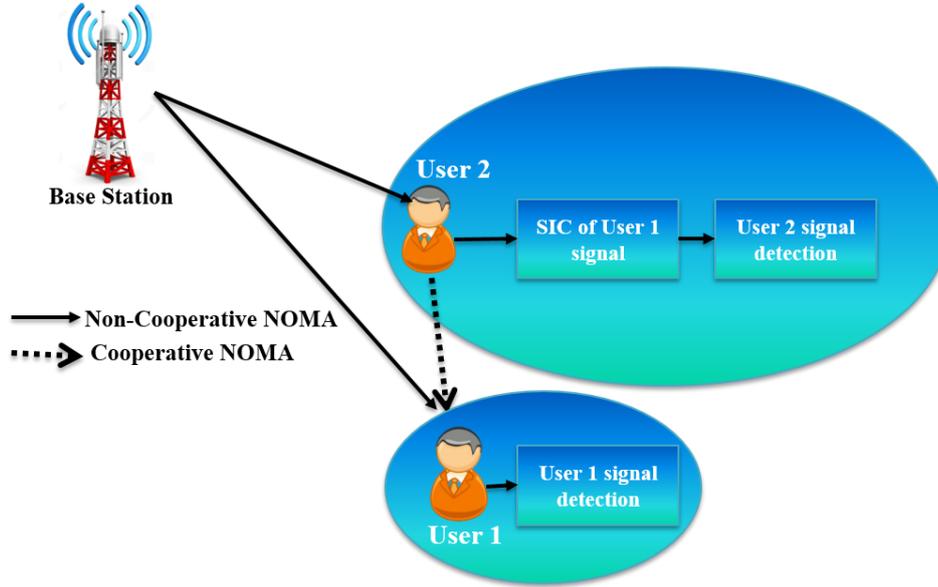

**Figure 5.** Cooperative NOMA communication.

**Table 4.** Summary of studies on CC and NOMA

| Reference | Category | Transmission | Key Potential |
|---|---|---|---|
| [14] | Relay assisted NOMA | FD-DF (Uplink and Downlink) | Focuses to enhance maxmin sum rate of uplink and downlink rates |
| [37] | Relay assisted NOMA | FD-DF (Uplink and Downlink) | Superior sum rate performance to HD/FD OMA and HD-NOMA |
| [52], [80] | Cooperative NOMA | Downlink | Reduced system redundancy, improved diversity gain and user fairness. |
| [81], [82] | Relay-assisted NOMA | Uplink/Downlink | Network coverage extension |
| [83], [84], [85] | Multi-cell cooperative NOMA | Uplink/Downlink | Improves spectral efficiency and improves performance of cell-edge users. |
| [86] | Relay assisted NOMA | HD-DF (Uplink and Downlink) | Better maximum sum rate |
| [87] | Relay assisted NOMA | HD-AF and HD-DF (Uplink and Downlink) | Better maxmin user rate |
| [88] | User assisted NOMA | HD-DF (Uplink and Downlink) | Maximum ergodic sum rate, low complexity |

*4.2. VLC-NOMA*

Although VLC has a number of noteworthy benefits, but it also has some significant shortcomings. When it comes to the design of high-data-rate VLC systems, the comparatively constrained modulation bandwidth of LED is a major challenge. While typical OMA techniques do not encourage effective resource consumption, VLC allows multiple users to connect to the network. NOMA, and specifically PD-NOMA, can be seen as an appropriate multiple access approach that offers enough bandwidth for indoor VLC systems [89]. Significant gains in throughput and other performance metrics are possible with NOMA. Hence, NOMA can be thought of as a strong and effective multiple-access technique for increasing the spectrum efficiency of VLC systems. A small number of users can be efficiently superimposed by NOMA and VLC systems, which use LEDs as transmitters. The LED sources serve as entry points for a certain group of users in enclosed locations. The SIC process of NOMA requires CSI at both the transmitter and receiver sides in order to support user power allocation, de-multiplexing, and decoding order. This problem does not occur as frequently in VLC systems since the channel is comparatively steady. As a result of the LED and PD's near proximity in VLC networks, NOMA performs better when the SNR is high. By adjusting the PDs' and LEDs' FOVs, channel gain differences between users in VLC systems can be reduced. NOMA, which performs well even when user channels have significant variances in gain, is the best multiple-access strategy for very low latency VLC systems [90]. NOMA schemes outperform the present OMA scheme in terms of performance for VLC systems. Because of the rise in SIC computational complexity brought on by a growth in user count, it is difficult to continuously deploy NOMA to all users. User pairing is used as an alternate strategy to address this problem and reduce the complexity of the SIC decoding. The performance of NOMA-VLC systems can also be enhanced by compatible power allocation and user pairing [91]. Figure 6 presents an overview of downlink NOMA-VLC system for two users. Table 5 summarizes recent works on NOMA-based VLC.

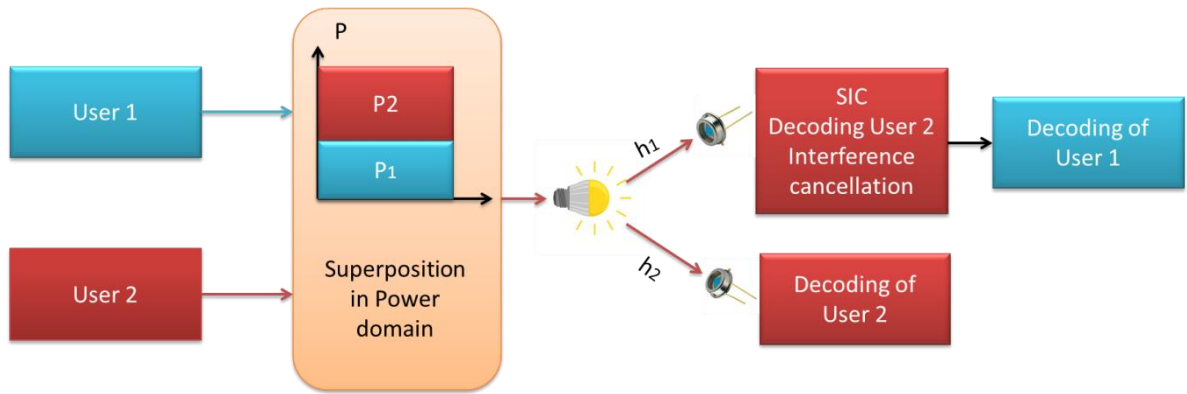

**Figure 6.** An illustration of downlink NOMA-VLC.

**Table 5.** Summary of recent studies on NOMA-VLC. [56]

| Reference | Communication Domain | Research Contribution |
|---|---|---|
| [24] | VLC | Review of potential challenges and future aspects for VLC-NOMA |
| [25] | VLC | Integration of emerging technologies, issues and future aspects of VLC-based NOMA |
| [56] | VLC | GRPA method to improve NOMA performance |
| [92] | VLC | Karush–Kuhn–Tucker (KKT) optimality conditions to enhance sum-rate performance than OFDMA |
| [93] | VLC | SCA techniques to achieve better sum-rate performance for optimized VLC-NOMA |
| [94] | VLC | Standard interior point method to attain higher sum-rate performance than FPA and GRPA |
| [95] | VLC | Iterative algorithm to attain an additional sum-rate gain of 10 Mbps for NOMA users than TDMA |
| [96] | VLC | Harris Hawks optimization to enhance sum-rate performance by jointly optimizing PA and UAV's placement |
| [97] | VLC | Gradient projection algorithm (GPA) to gain higher sum-rate performance than OMA |

| [98]  | VLC                       | Interior point approach and ZF-pre-coding NOMA technique outperform traditional NOMA and ZF |
| [99]  | VLC                       | EPA technique to improve energy efficiency of NOMA-based IoT networks |
| [100] | VLC                       | NGDPA method to enhance MIMO-NOMA's capacity |
| [101] | VLC                       | Power allocation and user pairing approaches for downlink VLC-based NOMA |
| [102] | OFDM-based VLC            | Analytical method and PA method for better maximum sum-rate than FPA and GRPA |
| [103] | Power-line fed VLC        | KKT optimality conditions to attain higher system sum-rate than FPA and NGDPA |
| [104] | Hybrid OMA and NOMA VLC   | Dynamic programming based layer-recursion model for better achievable throughput than traditional NOMA and TDMA |
| [105] | Hybrid RF/VLC             | Link selection and user pairing in Co-NOMA |
| [106] | Hybrid RF/VLC             | Outage performance and reliability improvement of Co-NOMA |

*4.3. UAV-NOMA Assisted Communications*

The combination of UAV with NOMA is an emerging field with many unexplored research avenues. The majority of rate optimization techniques in use today operate with one UAV operating as a flying BS and numerous users. This model was expanded to take into account the interference experienced by consumers served by a ground BS in [107]. However, the UAV-assisted NOMA system includes a variety of ways to incorporate the UAV, as explained in [108], and sum-rate optimization problems in these frameworks are attractive areas for further study. As an illustration, the use case involving the UAV as a flying user is explored in [109]. The introduction of NOMA to a multi-tier terrestrial and aerial framework involving ground BSs for long-term deployment, and certain HAP UAV BSs for medium-term deployment have been discussed in details in [108]. Future studies should focus on how to solve the sum-rate optimization problem when NOMA is incorporated into such complicated system models. Furthermore, with the added flexibility of the UAV altitude and placement, the rate optimization algorithms from other parts examined in this study can be applied to a UAV-NOMA system. Systems like MIMO and mMIMO can be incorporated with UAV-assisted NOMA systems for enhanced performance. Several antennas on the ground BSs, UAVs, or users might cause issues with beam design, power allocation, UAV positioning, and user schedule optimization. Similar to how relay-based cooperative methods can be modified to use the UAV as a relay. As a result, there are more options available because the relay can be placed closer to the BS or the user as needed. The UAV might even switch between the transmissions nodes when acting as a relay link. This results in some intriguing design issues with power allocation coefficients. Since the BS and UAV can work together to jointly establish the design variables in a UAV-NOMA system that attempts to maximize the sum rate, CoMP-NOMA systems with UAV represent another significant untapped field. The UAV-NOMA model can also be used with other B5G technologies; for instance, the authors in [110] used it with self-sustaining backscatter networks that does not need an external power source [111]. Table 6 discusses some studies reported on UAV-NOMA assisted communication systems. Figure 7 presents as overview of UAV-assisted NOMA for ground BS-served users and UAV-served users.

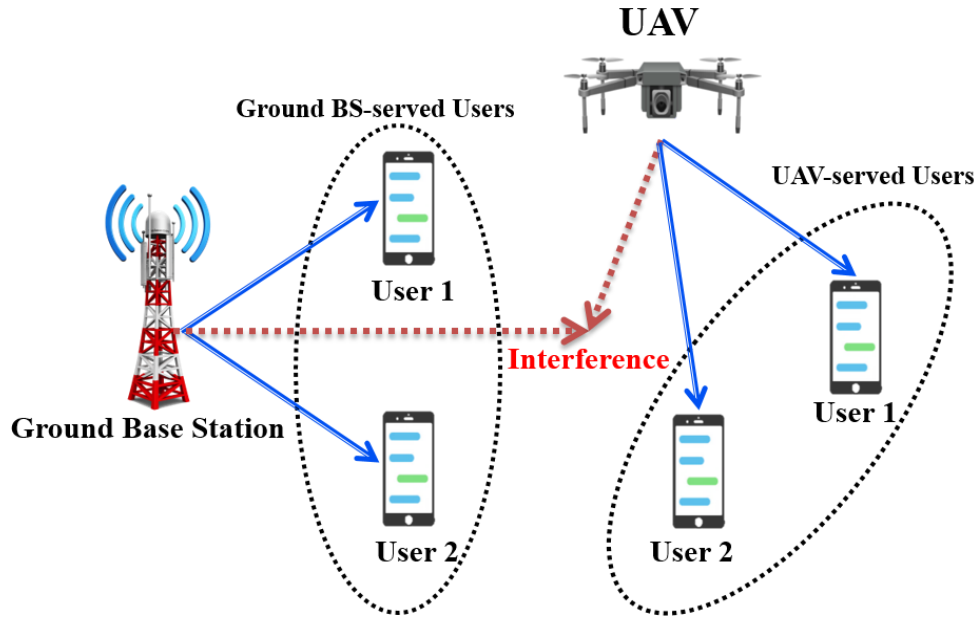

**Figure 7.** An overview of UAV-assisted NOMA

**Table 6.** Summary of studies on UAV-NOMA assisted communication systems

| Reference | System model | Design objectives | Optimization technique | Key findings |
|---|---|---|---|---|
| [107] | Multiple users, one UAV, one BS | Maximum sum rate | Iterative algorithm through BCD technique | Low complexity algorithm can steer UAV away from BS served users and close to its users |
| [109] | Ground users, UAV as a user and one BS | Maximum WSR of users and UAV | SCA technique and alternating algorithm | It achieves better sum-rate gains than non-cooperative schemes and OMA |
| [112] | Multiple users, one UAV, multi-antenna BS | Maxmin user rate | Penalty dual decomposition technique | Proposed techniques outperform OMA |
| [113] | Multiple users, one UAV | Maxmin user rate | Path following approach | Join optimization of various variables improves the sum rate than OMA |
| [114] | Multiple users, one UAV | Maxmin user rate | BCD method | Join optimization of user scheduling and UAV placement can double the sum rate than TDMA |
| [115] | Multiple users, one UAV | Maximum sum rate | Convex problems with KKT optimality conditions | The proposed power allocation and UAV placement approach outperforms fixed PA NOMA and OMA |
| [116] | Two users, one UAV | Maximum sum rate | Exhaustive search | The given power allocation and UAV altitude approach performs better OMA |

*4.4 Backscatter Communications (BackCom)*

Future wireless networks will benefit from ongoing exciting research in rate optimization strategies in BackCom networks. It is necessary to take into account system models with numerous backscatter receivers or RF sources, similar to those used in conventional multi-cell cellular networks. The source carrier wave on which the BDs will modulate their information must be selected. Also, according to [117], OMA performs better than NOMA for BDs using the harvest-then-transmit (HTT) scenario. When the BDs are merely sending on top of the RF signals emitted from the source, NOMA has been demonstrated to improve the sum rate. Future study could therefore focus on hybrid systems where certain BDs are in HTT mode and others are just perform modulating on top of carrier signals. In a recent study [118], the authors proposed an optimization strategy based on KKT conditions and dual theory to enhance the energy efficiency of backscatter-assisted NOMA vehicular networks.

*4.5. NOMA in RIS-Enhanced Networks*

Reconfigurable intelligent surfaces (RISs) have the potential to improve received SINR by creating a software-defined wireless environment by smartly changing the wireless propagation environment using passive reflecting elements [119]. NOMA-assisted RIS networks are capable of greater spectral efficiency and improved massive connectivity than OMA-assisted RIS networks. Most crucially, NOMA-assisted RIS networks are more adaptable than traditional PD-NOMA-based networks because RISs can modify the phase shifts of reflecting components and their placements to enhance the channel quality of specific users. Smartly controlling these phase shifts regulates the channel conditions for the network users. The severe restriction on the number of antennas at the transmitters and receivers can also be alleviated by integrating RISs in MIMO-RIS networks because RISs have numerous reflecting components. Traditional CSI-based SIC increases implementation complexity for RIS-enabled networks because error-free CSI for cascaded channels at the transmitter is introduced. Networks can manage the trade-off between channel estimation complexity and good spectral efficiency by integrating hybrid-SIC into NOMA-RIS [120]. So, by simultaneously increasing the spectral efficiency of RIS-enhanced networks and the design flexibility of NOMA-based networks, RIS integration into NOMA-based networks helps both RIS-enhanced wireless networks and NOMA-based networks. Table 7 present a comparison of studies on IRS-assisted NOMA systems.

Table 7. Comparison of works on IRS-aided NOMA [121].

| Context | [122] | [123] | [124] | [125] | [126] | [127] | [128] | [129] |
|---|---|---|---|---|---|---|---|---|
| Line-of-sight (LoS) consideration | ☑ | ☑ | ☑ | ☑ | - | - | - | Yes |
| Non LoS (NLoS) consideration | - | - | - | - | ☑ | ☑ | ☑ | - |
| Hardware impairment | - | - | - | - | ☑ | ☑ | | - |
| Op analysis | - | ☑ | - | ☑ | ☑ | ☑ | ☑ | - |
| Achievable rate optimization | ☑ | - | ☑ | - | - | - | ☑ | ☑ |
| OMA comparison | ☑ | ☑ | ☑ | ☑ | ☑ | - | - | - |
| NOMA system | ☑ | ☑ | ☑ | ☑ | ☑ | ☑ | ☑ | ☑ |

*4.6 Edge Caching and Mobile Edge Computing (MEC)*

The integration of NOMA with MEC networks is now possible due to the exponential rise of IoT devices with limited batteries in both current and upcoming wireless networks. In particular, the NOMA scheme can enable massive connections and improve the spectral efficiency of such networks to ensure effective communication. A resource allocation issue that maximizes the weighted sum rate of NOMA users in a downlink NOMA-enabled MEC network has been put forth in [130]. Subsequently, [131] has evaluated an adaptation of an earlier study [130] that jointly optimizes the resource allocation and power allocation. Future research would be interesting by examining the joint user clustering, resource distribution, and power distribution issue for downlink and uplink NOMA-enabled MEC networks. Similarly, NOMA assisted caching enables local content servers to be updated during periods of high demand using push-then-deliver and push-and-deliver techniques [132]. In [132], the cache hit probability served as the primary performance requirement. Further works in this area should focus on optimizing NOMA power allocation coefficients to boost system data rate and hence improve user fairness. In a recent study [133], the authors have focused on security issues for NOMA based UAV-MEC network. NOMA-MEC scenario for both uplink and downlink is shown in Figure 8. Table 8 summarizes some studies on MEC-assisted NOMA.

Table 8. Summary of reported works on MEC scenario [134], [19]

| Reference | NOMA Integration | Key Potential |
|---|---|---|
| [130] | Yes | The given NOMA scheme give better SWR performance as compared to OMA |
| [131] | Yes | The proposed optimal NOMA gives better WSR performance than OMA and |

| | | proposed scheme in [130] |
|---|---|---|
| [134] | Yes | It fastens convergence for better solutions than traditional optimization methods |
| [135] | Yes | Better average sum-rate performance than TDMA and NOMA without cooperation |
| [136] | Yes | The proposed cache-based NOMA outperforms NOMA without cache and cache-based OMA in terms of file delivery time and sum rate |
| [137] | Yes | Better performance than binary computation offloading mode, and NOMA performs better than TDMA regarding computation efficiency |
| [138] | Yes | It maximizes the number of users for task offloading and decreases average offloading delay |
| [139] | Yes | The DRL-aided approach can attain the near-optimal offloading with sufficient learning |
| [140] | Yes | Improved performance than FDMA-aided MA-MEC |

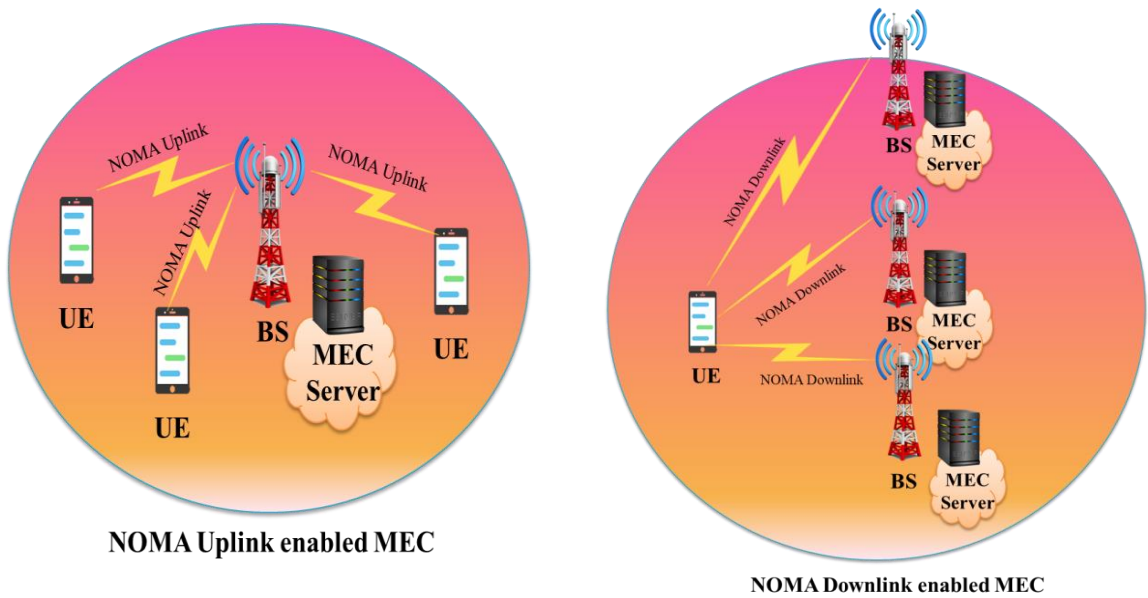

**Figure 8.** An overview of NOMA-MEC for uplink and downlink.

*4.7. Extension to MIMO*

To further increase the achievable spectral efficiency by utilizing the multiplexing gain and spatial diversity gain of MIMO systems, it is needed to modify the current NOMA techniques to their MIMO-assisted counterparts, particularly to large-scale MIMO systems. However, MIMO-aided NOMA technique is not simple in design aspects. Considering PD-NOMA as an illustration, where primary principle is to distribute the transmission power to users in an inverse relationship to the channel conditions. As channel gains and attenuations are scalars, it is possible to compare the channel characteristics of users for situations involving single-antenna nodes. But, in MIMO settings, a matrix is used to represent the channels. Hence, choosing which user's channel is superior becomes challenging. NOMA-solutions experience implementation issues as a result of this challenging task. Due to limited available solutions in the literature, it is still an open research topic. One option is to create numerous beams at the BS, where MIMO precoding and detection is utilized to reduce inter-beam interference and NOMA techniques are employed to assist users served by the same directional beam [141]. In such cases, the NOMA power allocation limitation must be kept into account for the beamforming design in order to allocate different beams to different users independently [142]. Table 9 discusses some studies reported on MIMO-based NOMA. An illustration of NIMO-NOMA system is presented in Figure 9.

**Table 9.** Summary of studies of NOMA based on MIMO [143]

| NOMA Scheme | Metric | Key focus | Challenges | Advantages |
|---|---|---|---|---|
| SA-NOMA [49] | OP | Inter-cluster interference | Imperfect CSI and SIC | High diversity gain |
| C-BF-NOMA [84] | Th | Inter-cell, inter-cluster interference | Imperfect SIC | Cell-edge users' throughput enhancement |
| ZFBF-NOMA [142] | Th, SE | Inter-cluster interference | Multicell case | Improves overall throughput |
| H-NOMA [144] | OP | Transmission power | - | Decreased the transmission power |
| PH-NOMA [145] | SR, OP | Intra and inter-cluster interference | MIMO-NOMA case | Reduced total power consumption |
| NOMA-SM [146] | EE, SE | Inter-user interference | Imperfect CSI | Improved SE |
| NOMA-MRT [147] | SR | Sum rate | Extension of mMIMO | Maximizes weighted sum-rate |
| ROBUST-BF-NOMA [148] | SR | Inter-beam, inter-cluster interference | Extension of MU-MIMO | Maximizes worst case sum rate |
| Random-BF-NOMA [149] | Th | Inter-cluster, inter-beam interference | Imperfect CSI | Decreases CSI feedback |
| NOMA-BF [150] | SR | Inter-cluster, inter-user interference | Imperfect CSI | Enhances QoS |
| NOMA-HARQ [151] | SE | Unreliable MCS adoption | Addition of MU-MIMO | - |
| PD-NOMA-SSK [152] | SE, BER | Secrecy | Power allocation | Enhances network throughput |
| NOMA-GSSK [153] | SE, BER, EE | Low SE of cell-edge users | SIC | Decreases computational complexity |
| NOMA-SSK [154] | SE, BER, EE | Low SE of cell-edge users | Power allocation | Decreases decoding complexity |
| TAS-NOMA [155] | SR | Multiple transmitting antennas | Imperfect CSI | Reduces, power, complexity and cost |

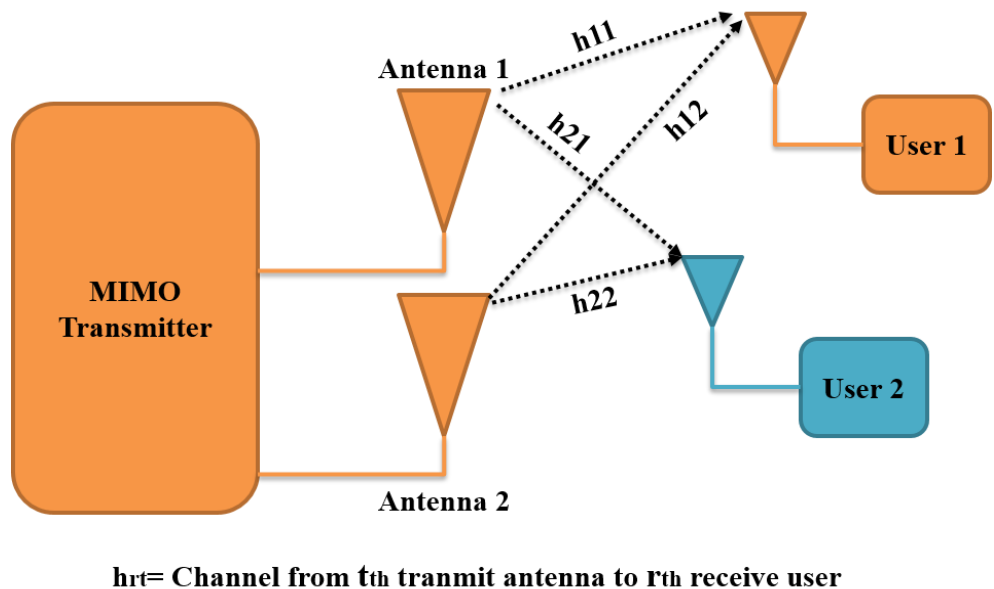

$h_{rt}$= Channel from $t_{th}$ tranmit antenna to $r_{th}$ receive user

**Figure 9.** An overview of MIMO-NOMA system.

*4.8. Cognitive Radio Inspired NOMA*

Using the idea of cognitive radio (CR) networks makes it easy to demonstrate the benefit of NOMA approaches [156]. In the context of CR networks, the user empowered with poor channel characteristics in a NOMA system might be seen as the primary user. If conventional OMA is utilized, even though the primary user is having a poor connectivity to the BS, no other users can access the bandwidth given to this user. Because the frequency or time slots are fully assigned to this user. The advantage of NOMA is to serve additional secondary users through particular beam that the prime user is occupying. Although the primary user's performance may suffer as a result of these secondary users, the system's overall throughput can be greatly increased, especially if the secondary users have good connectivity to the BS. The performance advantage of NOMA over traditional OMA may be easily demonstrated using the attractive idea of cognitive radio networks, greatly simplifying the demonstration of NOMA systems [141]. For instance, the method of optimal power allocation is tough in MIMO scheme conditions or when co-channel interference is present since it is difficult to determine the quality-order of the users' channel characteristics. The use of CR networks can place additional restrictions on power distribution, which must balance throughput and fairness similarly to conventional NOMA. Table 10 summarizes studies reported on CR-enabled NOMA. A combination of CR with NOMA is illustrated in Figure 10(a) and 10(b).

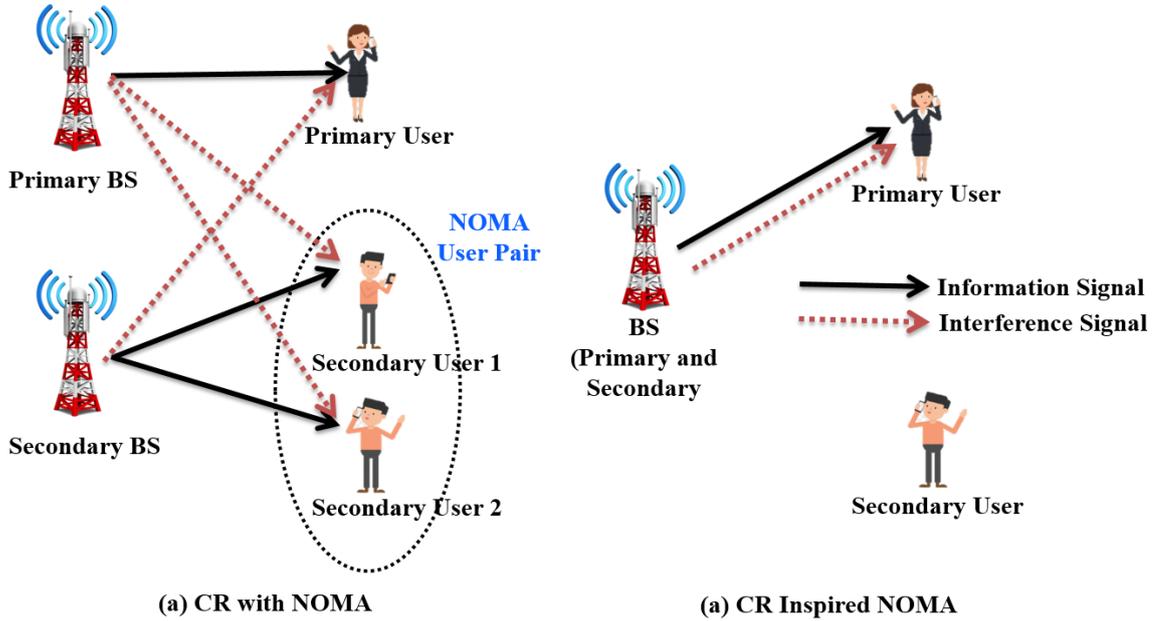

**Figure 10.** Combination of CR and NOMA (a) CR with NOMA, (b) CR Inspired NOMA

**Table 10.** Summary of studies of NOMA with CR [19], [23]

| Reference | Year | Algorithm | Objective |
|---|---|---|---|
| [76] | 2018 | Linear search algorithm | Maximum secondary transmitter rate for joint NOMA and OMA |
| [157] | 2018 | Lagrangian dual method, machine theory, bisection algorithm | Outperform traditional Cognitive OFDMA considering maximum weighted sum rate of secondary network |
| [158] | 2018 | Heuristic algorithm, dual decomposition method, binary search, SCA technique | Achieve secondary users fairness and enhance minimum video transmission quality |
| [159] | 2018 | Optimization problem using reformulation | Reduce the transmit power while attaining same rate performance results |
| [160] | 2018 | SDR along with line-search method | Enhance the user rate of both near and far users |
| [161] | 2019 | KKT optimality and D.C. programming transformation | Better sum rate performance for secondary-tier small cell network |
| [162] | 2019 | Greedy algorithm, KKT optimality, SCA techniques | Improve the sum rate of femto cell users |
| [163] | 2019 | Dichotomy method | Optimal sensing time is used to attain a maximum sum rate |
| [164] | 2019 | Heuristic algorithm | To attain minimum interference |
| [165] | 2019 | Monte carlo simulations | Evaluation of interference reduction |
| [166] | 2019 | Heuristic algorithm with various frameworks | To overcome channel modeling problems |
| [167] | 2020 | Monte carlo simulations | Error propagation analysis in SIC implementation |
| [168] | 2020 | Dual decomposition method and SDR approach | Better sum rate performance than OMA |
| [169] | 2021 | Heuristic algorithm | Maximization of PFEE taking uncertainty in the battery energy and channel gains |
| [170] | 2022 | Monte carlo simulations | Analysis of ergodic sum rate (ESR) and outage probability (OP) |
| [171] | 2023 | Monte carlo simulations | Performance analysis of classification error, packet error rate, and throughput |

*4.9. Deep Learning Inspired NOMA*

Deep learning-aided NOMA systems have recently been used in a number of different applications. A system can learn and develop using "deep learning" as opposed to using pre-established rules. It has various advantages over standard machine learning techniques, including the ability to work with enormous volumes of data available from complex networks for analysis purposes due to the expansion of network sizes and usages. With the computing capacity of graphics processing units, these metadata profiles are capable of producing end-to-end categorization solutions. An effectively trained system can sense any data provided to it, extracting the pertinent information and using that understanding to find and fix problems. Reinforcement learning (RL), supervised learning, and unsupervised learning are different kinds of DL techniques [172]. A system learns to carry out decisions on the basis of samples that have already been classified in supervised learning. Applications for supervised learning include regression and classification. Unsupervised learning reveals latent structure in an input and enables a machine to make decisions using unlabeled data. Algorithms based on unsupervised learning are frequently used to solve clustering problems. A system's capacity to learn new skills through repeated practice is referred as RL. RL does not need any input data to work because it learns from its environment. Signal detection and classification may be done entirely automatically due to reinforcement learning. DL methods are implemented using neural networks [173]. The three parts of a straightforward neural network are represented by the hidden layer, input layer, and output layer. In deep neural networks (DNNs), there are multiple hidden layers between the input and output levels. Each layer of the network's central processing units are neurons. Convolutional neural networks (CNN) and recurrent neural networks (RNN) are two forms of DNNs that have different methods for handling inputs and producing results. The authors discussed DL-related concerns in a NOMA-based system in [173].

The review papers [69], [174–175] provide a comprehensive discussion of DL-assisted NOMA systems. In [16], the authors compare DL-assisted NOMA systems and integration of several emerging technologies. They describe the opportunities, challenges, and future aspects for DL in NOMA systems. They demonstrate how the system capacity, channel state estimation, and spectrum efficiency can be improved using the DL-assisted NOMA system. The contribution of NOMA in the communication system is highlighted in [18]. The complete advantages of NOMA combined with various technologies, such as mobile edge computing (MEC), cognitive radio (CR), physical layer security (PLS), massive MIMO, visible light communication, and energy harvesting are offered. In [20], a complete analysis of downlink NOMA is provided while keeping networks into consideration using orthogonal multiple access (OMA) techniques with two to k users per cell. It discusses the major elements of NOMA that increase the spectrum efficiency of the system. Many elements and challenges of the NOMA system, including power allocation, CSI, inter-channel channel interference, and SIC factors, are discussed in details. Also, the effectiveness of deep learning (DL) and machine learning (ML) for the NOMA system is investigated in [176]. The paper discusses the potential of DL and provides a full examination of the problems with the wireless system. The distribution of resources, signal decoding, and the signal constellations are all covered. The authors in [177] provides in-depth discussion of NOMA and recent cellular and IoT network. The authors discuss the challenges associated with user clustering, channel access for densely coupled devices, interference mitigation, and network coverage expansion. They also covered the fundamentals of DL techniques and how the DL algorithm is used for resource management of IoT-enabled devices. Some recent studies have used DL approaches to improve the NOMA system's functionality. Figure 11 present an illustration of NOMA-based VLC system with a DL-aided signal demodulator to successfully attenuate linear and nonlinear fluctuations. A summary of recent studies on DL-aided NOMA is provided in Table 11.

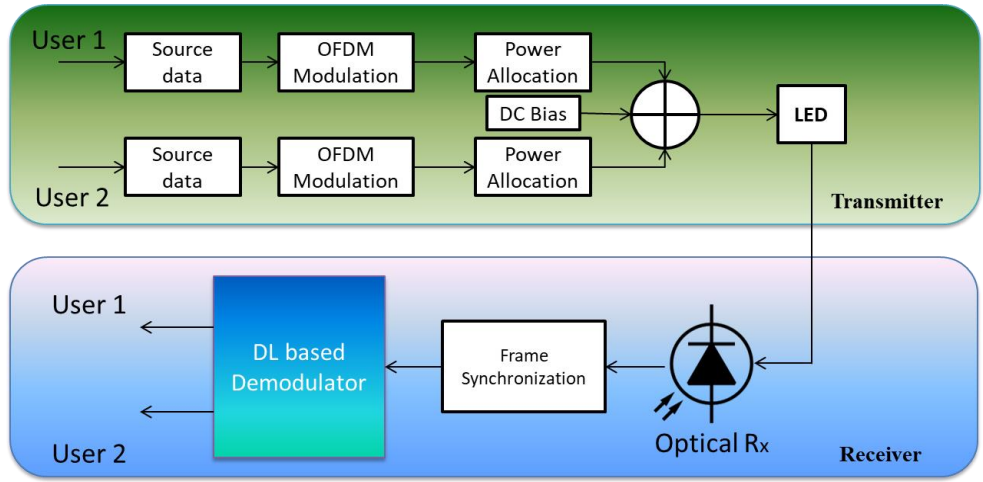

**Figure 11.** An overview of DL-aided NOMA-VLC system.

**Table 11.** Summary of recent reviews on DL-based NOMA.

| Reference | Focus Area | Research Contribution |
|---|---|---|
| [16] | DL-based NOMA | It discusses key performance indicators, integration of emerging technologies, challenges, and future aspects for DL-empowered NOMA systems. |
| [18] | Benefits of NOMA | It highlights the advantages of NOMA and the incorporation of MIMO, CR and MEC. |
| [20] | Discussion of NOMA | It addresses NOMA's influence on multi-cell networks. |
| [27] | DL-based NOMA for 5G networks | It outlines DL-based NOMA systems for resource allocation, power allocation, signal detection, user detection along with challenges and benefits of DL methods for NOMA. |
| [48] | NOMA in the context of next-generation multiple access | It addresses NOMA for 6G networks, future trends and research opportunities. |
| [71] | DL-based power allocation in NOMA | It discusses power allocation optimization using DL techniques and future research trends. |
| [173] | Review of DL-aided NOMA | It reviews DL-aided NOMA techniques, integration with various technologies, challenges, advantages, and future research directions. |
| [176] | Analysis of DL-aided NOMA performance | It examines DL-aided NOMA, advantages and its challenges. |
| [178] | Impact of DL on wireless communication | It focuses on sum rate maximization for DL-based NOMA. |
| [179] | DL for downlink MIMO-NOMA | It presents a survey of DL in SP blocks of downlink MIMO-NOMA systems and future trends. |
| [180] | DL-empowered NOMA transceiver designs for massive MTC | It highlights challenges and future trends for DL-empowered NOMA systems in the perspective of future mMTC. |

*4.10. HetNets and NOMA*

The network traffic is expanding because of proliferation of smart devices. This strain on network capacity requires spectrally efficient technologies. HetNets [181] is an emerging technology based on high-powered macro cell infrastructure containing low-powered small cells. This cell densification strategy offers several advantages including enhanced connectivity, improved QoS, enhanced network capacity, extended coverage [182], and efficient use of spectrum considering frequency reuse over smaller zones [183]. The integration of HetNet and NOMA is an effective strategy for 5G and B5G networks as both spectrally efficient technologies offer capacity enhancement. In this context, the idea of NOMA's user pairing can enhance the number of linked devices and further improve the throughput of system. A basic overview of NOMA-HetNet is presented in Figure 12 which contains macro BS and small BC to serve multiple users [1].

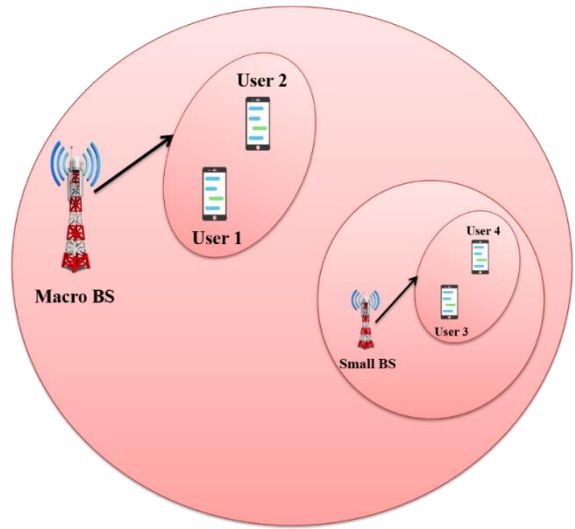

**Figure 12.** An illustration of NOMA HetNet

## 5. Applications of NOMA

This Section presents various application scenarios (see Figure 13) where NOMA techniques have proven their stature as discussed in existing research studies.

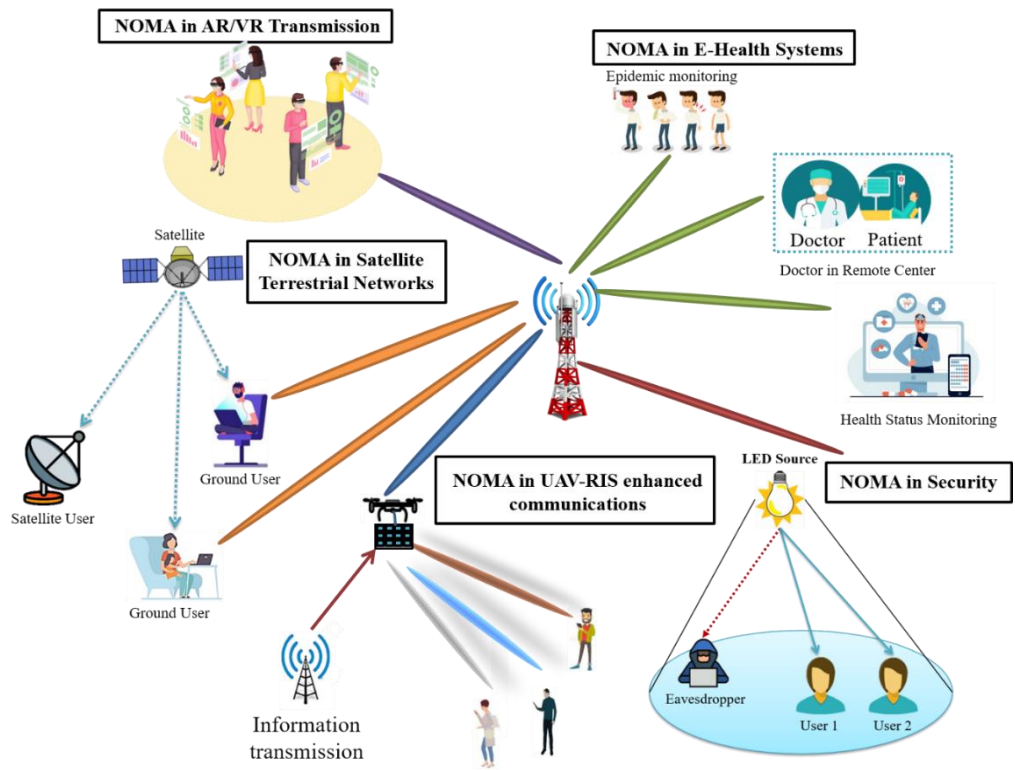

**Figure 13.** Applications of NOMA towards 5G

*5.1. NOMA in Integrated Terrestrial and Aerial Networks*

SAGINs, which fundamentally incorporate UAV-assisted terrestrial networks with satellites and other sky-platforms and, have gained attention in wireless communications research. The developed multi-tier SAGINs are able to fill the coverage gaps of single-tier networks by fusing the benefits of each network. This is accomplished by utilizing the integration of satellites, aircraft, drones, low-altitude platforms (LAPs), and high-altitude platforms (HAPs) [48]. In opposition to terrestrial wireless networks, the optimization difficulties confronted in SAGINs require massive coverage under several factors (power, BER, throughput, delay). For low-earth orbit networks, a limited number of satellites are required for complete coverage. In order to provide constant connectivity to everyone, everything, and everywhere, SAGINs depend upon the seamless incorporation of heterogeneous network elements. As hybrid SIC promises to address heterogeneous QoS demands in multi-tier SAGINs while ensuring widespread and huge coverage. The hybrid SIC-empowered NOMA may be used in SAGINs to improve performance of BER, latency and data rates.

*5.2. NOMA in Autonomous Networks/Robotic Communications [184]*

Wireless communications has captured a lot of attention with the integration of autonomous robots, which has significantly altered many socio-economic aspects of our society. Aerospace robots (such as UAVs, HAPs, airplane/airship), ground robotics (such as mobile robots, smart home robots, autonomous vehicles), and marine robots (such as underwater robots, unmanned submarine, unmanned ship) are different categories into which autonomous robots are divided. Prior to completely utilizing autonomous robotics, issues like collision tolerance and delay sensitivity must be resolved. Since, NOMA techniques are helpful to ensure low-latency and ultra-reliable communications, which can be integrated to meet the needs of autonomous robotic systems. Furthermore, because robots frequently work in dynamic environments and their channel conditions differ for each time slot, thus it increases the design flexibility for NOMA-aided networks. The received SINR of robots can be improved by dynamically placing them in the same cluster. Hence, NOMA can play a viable role to enhance the effectiveness and security of autonomous robotics.

*5.3. NOMA in E-Health Systems*

Recently, NOMA has garnered more interest for developing remote e-Health services because of offering massive connectivity and higher spectrum efficiency. The QoS-aided NOMA scheme can be used in communications-aware E-health systems. The QoS-based SIC system can be used to detect and decode the signal from the enquired user (delay-sensitive) and information from the monitored user (delay-tolerant), respectively. As a result, the limited communication resources can be used effectively. Hybrid-SIC empowered NOMA is useful for delivering huge connection with high QoS demands in E-health systems because there has been an exponential growth in the number of smart devices (such as smart glasses, smart watches, smart helmets) in communications-aware E-health systems. Moreover, it offers more access routes for shifting complex computations to edge servers or devices. Moreover, blockchain technology along with NOMA can be incorporated to secure medical record data of multiple patients.

*5.4. NOMA in VR/AR*

One of the important use cases for 6G networks has been identified as VR/AR, which provide customers a superior experience due to their immersive capabilities. In traditional OMA, it is difficult to break resources into smaller parts as they are distributed by blocks of a certain size, which implies OMA users must wait for ingress to resources. Since VR/AR video transmission networks are frequently delay-sensitive, which emphasize the significance of implementing NOMA in these networks [185].

*5.5. NOMA in WiFi Networks*

Although the idea of NOMA has been thoroughly studied for cellular systems, the future generation of WiFi systems can easily adopt it as well. Orthogonal resource allocation is still used in traditional WiFi networks. When all of the available orthogonal resource blocks are used by other users, it creates a challenging situation where some users cannot be served. NOMA technique allows for the simultaneous service for multiple users, which is crucial for the deployment of WiFi in congested environments like stadiums and airports [186]. Furthermore, distributed frameworks are required for the applicability of NOMA to Wi-Fi networks to support better service.

*5.6. NOMA in TV Broadcasting*

Another significant use of NOMA is in digital TV broadcasting [187]. It is worth noting that the NOMA concept has already been incorporated into the ATSC 3.0 next-generation digital TV standard, where it is known as Layered Division Multiplex (LDM). In particular, a TV station will superimpose multiple layers of video streams with various QoS needs and broadcast this combination to users. According to the channel characteristics for each user, specific layers of the video streams are decoded. It is also worth mentioning that upcoming wireless multicasting applications, a field that has not yet been extensively explored, may benefit significantly with the integration of NOMA into TV transmission.

*5.7. NOMA in Wireless Caching*

In order to effectively use wireless caching, content files must be actively pushed to local caching framework before being requested. Users can download these files through their local cache frameworks, even entirely bypassing the network controller. Previous research has shown that the NOMA concept increases the spectral efficiency of content distribution from the caching framework to the users as well as the timely and reliable forwarding of content files to local caching framework [132]. However, it is yet to be explored how to use NOMA in wireless caching networks to handle the dynamic variations in content distribution.

*5.8. NOMA in Internet of Things (IoT)*

The various traffic patterns of IoT devices are a major aspect of the IoT network. Especially, some devices require a lot of bandwidth, such as environmental monitoring cameras that transmit high-resolution photos or videos, whilst others require modest data rates but prompt delivery like intelligent transportation systems. By merging entities with different QoS needs at the same bandwidth, NOMA can be used to handle such situations. Additionally, recent research has shown that using NOMA in conjunction with encoding with finite block length might be a potential approach for enabling the Internet of Things. NOMA can play a vital role to ensure reliable communications while offering massive coverage for IoT networks [188].

*5.9. Underwater Communications*

Examining the sum-rate performance of an underwater acoustic NOMA system with the joint optimization of user clustering, resource allocation, and power allocation can be a promising subject for future breakthroughs in underwater communications networks. Another intriguing area for future research is the use of the rate-optimal NOMA system in multi-hop cooperative underwater optical wireless communications. The authors of [189] proposed an architecture for a software-defined hybrid underwater optical-acoustic network. This design aims to achieve low-latency and high-speed for ubiquitous network performance. Research on the implementing of NOMA technique with such an opto-acoustic system design could be could be a hot topic for future research. Table 12 provides a summary of some previously reported studies on NOMA with UWC.

Table 12. Summary of studies of NOMA with UWC

| Reference | Year | Algorithm | Objective |
| --- | --- | --- | --- |
| [190] | 2019 | Uplink PO-NOMA | Investigation of channel modeling |
| [191] | 2019 | Full-duplex DF relay method | Interference management and user fairness |
| [192] | 2019 | Heuristic algorithm | To investigate channel modeling |
| [193] | 2019 | Heuristic algorithm | To enhance the security |
| [194] | 2019 | Single receiver, diversity receiving method | To evaluate the interference management |
| [195] | 2021 | FDCR based NOMA | Evaluation of imperfect SIC and RSI |
| [196] | 2023 | KKT condition and the bisection method | Fairness of fair users and QoS of edge users |

**6. Potential Challenges and Open Research Problems**

NOMA can be used to overcome a variety of challenges in different use cases. However, implementing NOMA systems are prone to new obstacles and limits that necessitate proper solutions. On the basis of existing literature, in this section, we highlight several challenges and research problems that need to be explored in the future works. Figure 14 shows various challenges in NOMA.

*6.1. Hardware Complexity*

In comparison to OMA, NOMA results in more hardware complexity. The addition of the SIC detector to NOMA enhances the complexity of the electronics used to first distinguish signals with high and low power levels in order to gather user data. The employed SIC technique could result in significant limitation for battery if the number of users is particularly high or quick signal transmission is necessary [6]. Since consumer devices want their batteries to last longer, thus NOMA adoption may not be effective in ultra-dense networks. To solve this issue, efficient user clustering and power distribution are essential.

*6.2. Receiver Design*

For huge connectivity in 5G, the complexity of an SIC-aided receiver can reduce user performance. Consequently, a high-performance non-linear detection approach that is more precise and minimizes the influence of error propagation is needed to solve this problem. Although the MPA-based receiver is sophisticated, it effectively eliminates the problem of error propagation by employing the Gaussian distribution approximation method. When there are a lot of connections, as is anticipated in 5G, it produces better and more accurate outcomes. Additionally, by building a graph with check nodes, observation nodes, and variable nodes that match to the LDPC code's equations, the MPA can be utilized to jointly detect and decode the incoming symbols [15]. To enhance the effectiveness of signal detection, these receivers also communicate, demodulate, and decode the data symbols more effectively. To enhance performance at the receiver side, however, it is necessary to take advantage of issues like accurate signal recognition, error propagation, and efficient receiver design [197].

*6.3. Error Propagation in SIC Implementation*

The primary tenet of NOMA is to implement SIC detection on the receiving end to identify the user with the best channel characteristic. As a result, accurate assessment of the high power signal is necessary for accurate signal detection. SIC detection might be significantly impacted due to hardware and channel defects, which put an impact on the receiving process. Because of the presence of timing offset (TO), carrier frequency offset (CFO) and other hardware-based defects, optimal channel estimation is challenging for NOMA systems [141]. Particularly, low-quality clocks create substantial CFO and TO, which drastically lowers the quality of the transmission. These shortcomings could result in significant performance reduction even in these circumstances. Strong CFO and TO estimation are produced by the use of multicarrier waveforms such as OFDM. Hence, incorrect detection and error propagation in the SIC detection procedure can degrade NOMA's performance. Sophisticated solutions are required to fix this and enhance the transmission quality for NOMA users. Increasing the estimation accuracy of the aforementioned impairments is a more essential to enhance performance rather than modifying the primary detector components.

*6.4. Optimal Pilot Allocation*

When compared to OMA systems, intra-user interference and error propagation impact NOMA systems because several signals are transmitted in an overlapped manner. It is undeniable that a good performance requires a perfect or almost perfect CSI. Important design issues for the adoption of NOMA include pilot roles and the number of pilots allotted. Because of the unpredictable channel characteristics in wireless communication, these issues are crucial. However, because of the inherent interference, careful design and optimal pilot allocation are important for

NOMA systems. In order to allocate enough pilots at the right spots and achieve high performance of NOMA systems, channel characteristics should be tracked effectively and precisely.

*6.5. Instantaneous CSI Requirement*

Another fundamental problem of CSI estimation exists for NOMA implementations in addition to optimal pilot allocation. A severe issue arises when a hitherto allocated frequency band is given to a secondary user; the CSI for this user's transmission should be calculated using orthogonal transmissions. This inescapably disrupts the main user's transmission and creates an error propagation. Moreover, instantaneous band allocation in congested networks may be necessary as this problem could become more critical. The future NOMA systems should focus on an efficient and practical solution to overcome this issue. Optimal pilot allocation issues in massive MIMO systems may be taken into consideration as a possible solution [198].

*6.6. Channel Estimation*

For resource allocation or multi-user detection in majority existing works on NOMA, perfect CSI is taken into consideration. NOMA faces channel estimation errors since it is impractical to achieve perfect CSI in real-world applications. The impact of realistic imperfect channel estimation in NOMA systems and a low-complexity transmission rate back-off method was developed to reduce the effects of the channel estimation errors have been studied in [199]. Moreover, [61] and [200] explored the architecture of the practical channel estimation and optimization strategies for lowering the channel estimation error for NOMA. However, the rapidly expanding traffic of 5G network users will cause serious inter-user interference, which could lead to a serious channel estimation inaccuracy. Hence, novel and sophisticated channel estimation techniques in NOMA systems is a hot research topic for researchers.

*6.7. Grant-Free NOMA*

OMA has substantial transmission delay and signaling overhead due to the access grant-aided transmission used in downlink resource distribution and uplink scheduling. An MA approach that tackles grant-free transmission and offers low signaling overhead and low transmission delay is highly desirable. The aforementioned problems are handled by using NOMA as it will be able to enable vast connectivity, low transmission latency, little signaling overhead, and without the need for grant-free transfers. Hence, a contention-aided NOMA is a viable option for situations where contesting users are allotted one or more pre-configured resources. In contrast, integrated protocols, which include random backoff techniques, can be an alternative approach to resolve non-orthogonal collisions while lowering the rates of packet lost. Additionally, NOMA eliminates the reliance on accessing the grant method, which prevents BS from obtaining any information pertaining to the users. Due to the lack of user activity, compressed sensing (CS) based methods are therefore required to solve this problem [15].

*6.8. Resource Allocation*

NOMA is adopted to accurately distribute radio resources among users, enhancing user fairness, throughput, and data rates. Due to the restricted radio resources in the spectrum, resource distribution to users in a multi-cell situation becomes challenging as the number of user increases. Because of its potential to serve several users at once while utilizing various power levels, NOMA has can effectively allocate the resources. However, due to cross-channel and co-channel interference, allocating resources to users through the NOMA technique is rather difficult [201]. A suitable resource allocation strategy is needed to limit these interferences in order can reduce the influence of error propagation. In PD-NOMA, the power allocation method has a direct influence on the receiver capacity to mitigate interference. The BS can flexibly manage the cell-edge throughput, overall throughput and the users' fairness by precisely implementing power allocation. For achieving a close to optimal power allocation, dynamic programming or DL techniques may be taken into account. Moreover, dynamic power allocation is an interesting research area for future works.

*6.9. Enhancement in Physical Layer Technologies*

The pre-existing majority of NOMA techniques do not support an optimal design since these techniques only support symbol-level or bit-level operations. To solve this problem, a combined technique that uses both bit and symbol levels must be developed. However, it is susceptible to specific channel conditions and requires more practice to handle real-world scenarios. Apart from the design on the transmitter side, another physical layer technology [202] that requires improvement is the detection of signals on the receiver side. The existing approaches add delay by requiring a number of rounds to detect the symbol. In order to increase the system reliability, researchers should introduce novel methods to reduce latency and complexity of NOMA systems.

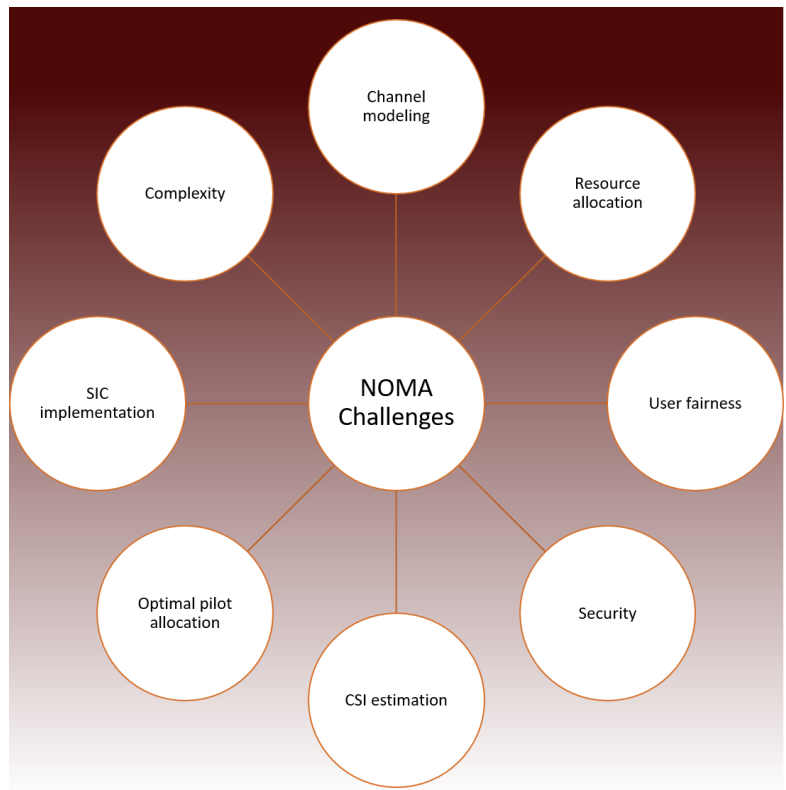

**Figure 14.** Basic challenges in NOMA

**7. Security Issues in NOMA**

Security is a crucial concern in every generation of wireless communication systems that must be resolved. It is important that the wireless transmission needs to be carefully monitored. In [203], the author put up a compelling idea for enhancing security of wireless channel through the use of cryptographic techniques at the physical layer. In contrary to [203], the researchers of [204] suggested a way to strengthen the network's confidentiality. Future wireless networks will include NOMA as a multiple access technique to support 5G and B5G use cases. However, NOMA security is crucial for the development and implementation of 5G wireless networks. The deployment of NOMA has some security issues that must be overcome, including SIC implementation transmit powers, accurate CSI, user security and privacy needs. Several studies reported on NOMA have focused on security issues. In particular, the capability of NOMA is that the power allocation coefficients are developed considering the channel characteristics of legitimate users, which ensures effective suppression of eavesdropping as SIC at the eavesdroppers might not be possible.

The authors of [205] and [206] investigate how to maximize secrecy sum-rate (SSR) in physical layer security of a NOMA network. The researchers in [205] develop a SISO NOMA network where the source must deliver data to each valid user at a given minimum data rate while maintaining a predefined QoS demand. It is noticed that there is a certain transmit power level that meets the minimal QoS criteria for all authorized users. The ideal transmit power allocation that optimizes SSR is determined by formulating an optimization problem, and the solution is validated by numerical outcomes. Using the random direction (RD) and random way point (RWP) models, the effects of a dynamic environment for two random mobile users are investigated under passive eavesdropping [207]. Under the restrictions of power and QoS, a rate maximization and sum average secrecy problem is formulated. A threshold power allocation approach is suggested to increase the average sum-rate and secrecy rate of mobile NOMA users. By using careful beamforming design and artificial noise (AN), the authors in [206] enhanced the PLS of a MISO-NOMA network. Finding the best beamforming parameters that can balance the SINR between the strength of the interfering signal and strength of the useful signal would improve secrecy performance. In [208], the secrecy outage probability (SOP) is taken into consideration as an optimization strategy to measure secrecy performance. The authors proposed a novel NOMA practical design for such situations where the CSI of the receiver is not entirely known at the transmitting side. The design characteristics carefully examine power allocation and decoding order factors. A strong near-legitimate user is combined with an attacker far weak user by the authors in [209] in order to investigate the PLS's performance in a NOMA system. In order to formulate an optimization problem, the outage probability of the NOMA pair for SOP and reliability of the legitimate user as a QoS performance parameter are established. Under the security constraint of optimal power allocation and SOP of the strong user, an optimization problem is formulated which reduces the pair outage probability. Outage probability and SOP closed-form formulas are generated. The authors of [210] evaluate the secrecy performance of a resource allocation problem by creating a secure and reliable resource allocation approach for a MISO MC-NOMA network. To maintain reliable communication in the existence of eavesdroppers, a robust optimization problem is formulated under QoS of legitimate users with AN and imperfect CSI of the eavesdropping channel. In [211], the authors described how to leverage stochastic configuration in a large-scale network with a single antenna for NOMA users to address the physical layer security issue. Using this method, the BS connects with the NOMA users which are randomly scattered. A secure zone is utilized around the BS to increase the physical layer's security while giving intended users more capacity than eavesdroppers. Furthermore, the authors of [212] demonstrated a multiple antenna-aided NOMA and suggested an approach to enhance physical layer security. Specifically, the authors of [213] considered NOMA principle to send unicast and multicast messages simultaneously. In addition, beamforming at the BS is smartly constructed to artificially enlarge the difference between the channel characteristics of two types of users, which can substantially enhance the secrecy data rate. Although the authors of [205] examined the security problem in SISO-NOMA networks, security of NOMA in the perspective of MIMO and mMIMO [214] is yet to be explored. The combination of PLS and NOMA is emerging research area, and further research contributions are required to design low complexity, practical and robust techniques to realize security and privacy in NOMA.

Some studies have focused on blockchain technology which guarantees the security of data being sent over NOMA as a multiplexed signal. Blockchain a distributed ledger technology where both public and private accesses are utilized. Using a public key to trace the user allows for the privacy of the hidden user. While a private key can be utilized to encrypt data so that it is protected. By employing blockchain technology, the problem of transmitting information to powerful users can be avoided. The NOMA technology requires more research into various cryptographic techniques in order to provide more useful and practical designs to secure NOMA transmission. Security issues and solution for NOMA have been summarized in Table 13. Figure 15 shows an overview of downlink NOMA based on near and far user in the existence of eavesdroppers.

**Table 13.** Summary of security issues and solutions in NOMA

| Reference | Solution method | Characteristics |
|---|---|---|
| [156] | Analytical | Overlay CR-NOMA, closed-form expression for EST, SOP, COP of PUs. Impact of transmit power and power allocation of secrecy performance of system |
| [205] | Optimization | Minimize data rate and optimal transmit power to ensure QoS of legitimate user, maximize SSR |
| [206] | Optimization | MISO-NOMA, optimal PA, beamforming with AN, Maximize SSR |
| [207] | Optimization | SISO, mobility models, passive eavesdropping, two random mobile users, maximize SSR, unknown CSI |
| [208] | Optimization | Minimum transmit power, CSI of eavesdropper is not perfectly knows, SOP |
| [209] | Optimization | Strong legitimate user and weak eavesdropper |
| [210] | Optimization | Secure and robust resource allocation, half-duplex users, full-duplex BS, imperfect CSI of eavesdropper |
| [211] | Analytical | Large scale NOMA network, single-antenna, eavesdropper exclusion area |
| [212] | Analytical | Large scale NOMA network, multiple-antennas, AN at BS, known CSI at BS |
| [213] | Analytical | Beamforming design, power allocation, and enhanced secrecy rate |
| [215] | Analytical | Enhanced secrecy rate through encoded data with IMEI and MAC address |
| [216] | Analytical | Users and BS in half-duplex, eavesdroppers in full-duplex |

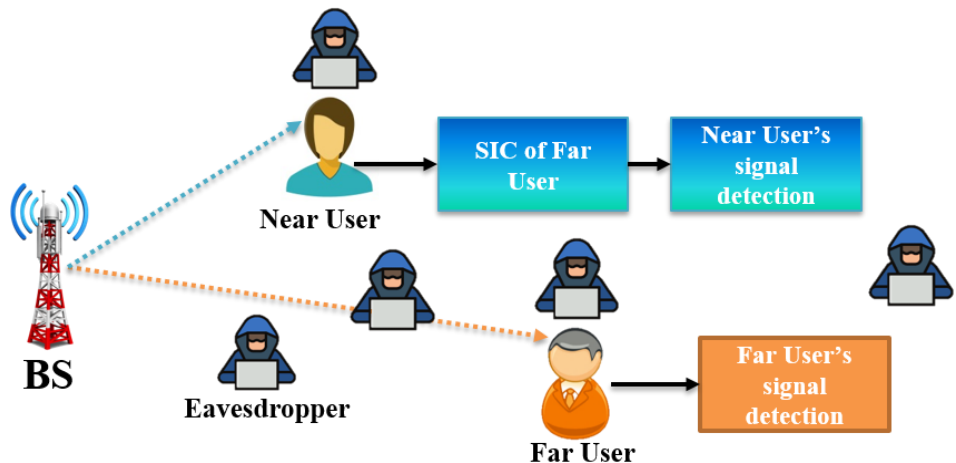

**Figure 15.** Downlink NOMA with near and far users and eavesdroppers.

## 8. Future Research Directions

On The basis of our learning from the existing studies in the literature, this Section focuses on several research directions that must be explored in the future works.

*8.1 Optimal/low-complexity receiver design*

Many issues of spectrum efficiency and energy efficiency can be resolved using optimization techniques in the NOMA systems. An ideal low-complexity receiver design can yield all of the advantages of NOMA. The receiver design in NOMA can take advantage of a message-passing mechanism. Yet, the main barrier to the massive connectivity is receiver's complexity. With SIC-aided receivers, the performance of some users might suffer from error propagation. Some studies have proposed unique receiver [217, 218], but a more thorough examination of the capacity gain is still required. Moreover, investigation of a reconfigurable receiver design is crucial for accurate detection to reduce complexity.

*8.2 NOMA with imperfect CSI*

The majority of NOMA research operates under the presumption of perfect CSI. Only a limited number of studies have addressed NOMA with imperfect CSI. For instance, the authors in [219] assess NOMA's performance with partial channel information. The authors in [220] present an improved CSI feedback mechanism for NOMA that has various benefits, such as user pairing and enhanced performance. The impact of imperfect CSI on various NOMA techniques must be examined, and a thorough throughput evaluation must be conducted. Hence, in NOMA systems, feedback techniques with fewer bits can provide reliable performance when there is substantial user interference. In order to obtain the required performance with less overhead, it is crucial to build a feedback method with the ideal number of bits.

*8.3 MIMO-NOMA and IoT*

One of the fundamental components of the upcoming wireless networks is the Internet of Things (IoT). High spectral efficiency and massive connectivity are required to meet the demand of expanding IoT devices. To overcome the computational complexity, it is crucial to propose joint beamforming and user allocation algorithms in MIMO-NOMA [221]. To address various QoS needs in IoT, MIMO-NOMA transmission techniques are highly desirable. A MIMO-NOMA system for IoT network has been proposed in [222]. Additionally, it is crucial to tackle issues like signaling overhead due to massive connectivity, resource allocation and receiver design.

*8.4 Optimal wireless power transfer*

An effective way to handle applications with limited energy is wireless power transfer. One of the main advantages of NOMA is that users close to the BS can have better channel characteristics than users farther from the BS [223]. To increase transmission reliability, users close to the BS can serve as relays for users located far away from the BS. Since the network devices have limited energy, wireless power transmission can be a viable solution. A few works have addressed the ideal design of NOMA with wireless power transfer (WPT) [224]. WPT in NOMA systems for IoT services is yet to be further explored. Investigation of NOMA with WPT along with user fairness and high data rate is also crucial. Hence, new optimized solutions for WPT in NOMA to improve spectrum and energy efficiency should be proposed. For energy harvesting in various NOMA systems, SWIPT is an appropriate contender. In SWIPT-NOMA, the strong user accumulates the energy from the signals transmitted from the BS and uses it to power the relay transmission. The researchers have not yet taken into account the circuit energy consumption, nonlinear energy harvesting properties, and hardware shortcomings. Future research studies can examine the influence of these characteristics on performance of SWIPT-NOMA. Figure 16 presents a basic design of NOMA-based SWIPT system.

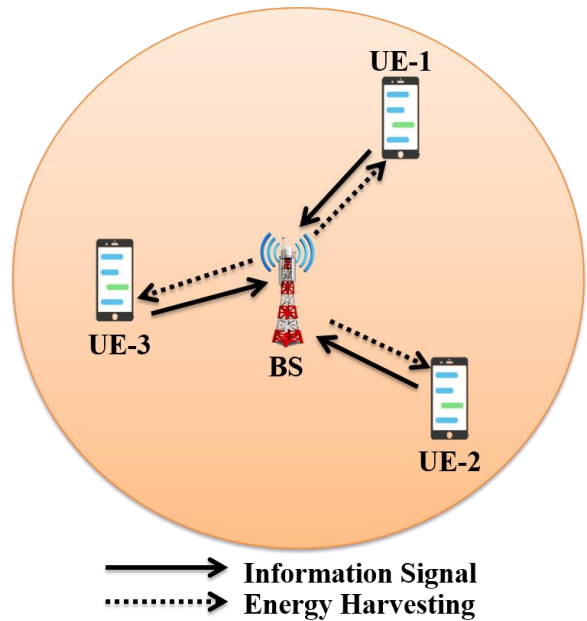

**Figure 16.** NOMA-enabled SWIPT.

*8.5. Joint Radar-Communication Systems*

When a single BS is used to carry out the communication and sensing operations utilizing the same frequency band, a joint radar-communication system may be able to alleviate the spectrum shortage problem. The beamformer will be tuned to meet the needs of both functions. NOMA can be used for simultaneous sensing, in addition to serving numerous users [225]. The joint radar-communication system based on NOMA are yet to be investigated to ensure the effectiveness of both sensing and communication components.

*8.6. THz and mmWave Communication*

At present, higher frequency bands like millimeter wave (mmWave) bands are being utilized in the 5G standard to meet the expansion in higher data demands. The mmWave spectrum, however, won't be sufficient to empower networks beyond 5G. Terahertz (THz) bands are wide open, and they can play an important role for developing future wireless networks [19]. However, the communication is difficult in scenarios with strict LoS conditions where several factors like shadowing and link misalignment can degrade the performance. Consequently, to better comprehend the real-world situations where NOMA performance would be optimal, it is therefore beneficial to examine the ideal pairings of NOMA users under specific error rate performance needs. Also, integration of technologies like IRS and cooperative communication can substantially enhance network coverage.

*8.7. Imperfections in OFDM-NOMA Systems*

The cyclic prefix (CP) length is considered to be longer as compared to the channel delay spread, while time or frequency synchronization are assumed perfect to completely eliminate inter-symbol interference in most of the studies reported on OFDM-NOMA systems. However, in real-world applications where inaccurate frequency and time synchronization can often occur, these ideal circumstances cannot be implemented. Furthermore, user mobility makes the OFDM-NOMA vulnerable to Doppler shift and subcarrier loss [226]. Consequently, there is still need to investigate OFDM-based NOMA system for better performance and reduced error rate.

*8.8. Hybrid-NOMA*

Hybrid NOMA is made by combining several emerging technologies with NOMA e.g., CDMA, FDMA, or TDMA to meet the future requirements of high data rate, massive connectivity, low complexity, and high energy efficiency. Several techniques have been proposed by scholars to define hybrid-NOMA. For MA, the hybrid-NOMA employs both CD-NOMA and PD-NOMA. Additionally, compared to NOMA and OMA approaches, hybrid-NOMA offers greater spectrum efficiency [227]. Moreover, NOMA and SDMA have been merged to increase user fairness and data rate [228]. Hybrid-NOMA has been developed by combining OMA and NOMA to increase network connectivity and spectrum efficiency.

*8.9. Spectrum sensing using NOMA*

Using high-performance spectrum sensing, NOMA can efficiently regulate interference in CRNs. Firstly, the performance analysis is accurate where NOMA approaches are not deployed in the core network, and traditional spectrum sensing approaches such as energy detection or eigenvalue-aided spectrum sensing are utilized. Secondly, the performance analysis are inaccurate where NOMA approaches are used for multiple users and the traditional spectrum sensing approaches are used in CRNs with NOMA. The difficult problem is determining the probability of detecting a correlation between samples. The deployment of non-orthogonality to design new spectrum sensing approaches is an open research issue.

*8.10. Mobility in NOMA in 5G/B5G*

The vehicular communication in 5G/B5G networks should keep NOMA users' mobility into consideration. It can result in a constant shift in user channel gains, and thus simulating the ideal user pairing for various numbers and types of devices which require access to a single RB and SIC. Because of mobility, the quasi-static NOMA channel must be reprocessed again in order to establish the optimal decoding sequence. Existing static models must incorporate new techniques to accommodate mobile NOMA environments. The dynamic user pairing approaches e.g., caching or using a compatible technique with a hybrid method may be utilized [23].

Figure 17 presents various future research aspects of NOMA systems.

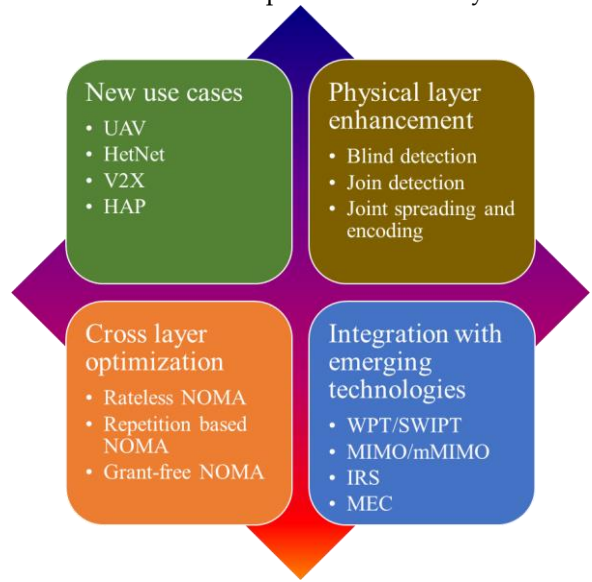

**Figure 17.** Future research aspects of NOMA

**9. Conclusion**

One of the essential promising technologies for completing the varied demands of 5G has been recognized as NOMA. NOMA outperforms OMA by granting several users permission to use the same radio resources. This survey provides a thorough analysis and discussion of the emerging

NOMA techniques. We started by outlining the basic ideas, guiding principles, and advantages related to the NOMA approaches. We also highlight the key factors to be taken into account while real-time implementation of NOMA systems. Moreover, the key performance indicators were in-depth reviewed in this survey article. We discussed several application scenario where NOMA systems can contribute to meet the ongoing expanding demands of wireless networks. We briefly explained the coexistence of NOMA with several emerging technologies and investigate the potential benefits of this incorporation. Furthermore, we outlined the NOMA technology's open challenges and security issues. Finally, we dedicated a section to future research directions. We anticipate that our survey will provide some guidelines for future research contributions.

**Funding:** This work is supported by the National Natural Science Foundation of China under grant 62261009, Ministry of Education Key Laboratory of Cognitive Radio and Information Processing (CRKL200106).

**Institutional Review Board Statement:** Not applicable.

**Informed Consent Statement:** Not applicable.

**Data Availability Statement:** Not applicable.

**Conflicts of Interest:** The authors declare no conflicts of interest. The authors declare that they have no known competing financial interests or personal relationships that could have appeared to influence the work reported in this paper.